\documentclass[journal,onecolumn]{IEEEtran}

\ifCLASSINFOpdf
\else
\fi
\usepackage{hyperref}
\usepackage{amsmath}
\usepackage{amssymb}
\usepackage{bm}
\usepackage{amssymb}
\usepackage{amsthm}
\usepackage{multirow,booktabs}
\usepackage{cases}
\usepackage{tabularx}
\usepackage{adjustbox}
\usepackage[figuresright]{rotating}
\usepackage{longtable}
\usepackage{booktabs}
\usepackage{multirow}
\usepackage{color}
\usepackage{cite}
\usepackage{multirow,booktabs}
\usepackage{cases}
\usepackage{tabularx}
\usepackage{adjustbox}
\usepackage[figuresright]{rotating}
\usepackage{longtable}
\usepackage{booktabs}
\usepackage{multirow}

\usepackage{enumerate}

\usepackage{lineno,hyperref,mathtools}
\usepackage{mathrsfs} 
\usepackage{threeparttable} 
\allowdisplaybreaks
  
\newtheorem{theorem}{Theorem}
\newtheorem{proposition}[theorem]{Proposition}
\newtheorem{remark}{Remark}
\newtheorem{problem}{Problem}
\newtheorem{claim}{Claim}
\newtheorem{definition}[theorem]{Definition}
\newtheorem{lemma}[theorem]{Lemma}
\newtheorem{corollary}[theorem]{Corollary}
\newtheorem{example}[theorem]{Example}

\definecolor{checkedcolor}{rgb}{0.5,0,0.5}

\usepackage{makecell}

\begin{document}
%
\title{
Optimal Ordering-Free Two-Dimensional Insdel Reed--Solomon Codes via Cross-Ratios
}
%
%
%
\author{Runqing~Qiu, Shixin~Zhu, Yang~Li, and San~Ling
\thanks{Runqing~Qiu and Shixin~Zhu are with the School of Mathematics, Hefei University of Technology, Hefei 230601, China (e-mail: runqingqiumath@163.com and zhushixinmath@hfut.edu.cn).}
\thanks{Yang Li and San Ling are with the School of Physical and Mathematical Sciences, Nanyang Technological University, 21 Nanyang Link, Singapore 637371, Singapore (e-mail: yanglimath@163.com and lingsan@ntu.edu.sg). San Ling is also with VinUniversity, Vinhomes Ocean Park, Gia Lam, Hanoi 100000, Vietnam (e-mail: ling.s@vinuni.edu.vn).
}

\thanks{This work was supported by Nanyang Technological University under Research Grant 04INS000047C230GRT01, the National Natural Science Foundation of China under Grant 12671649, and the China Scholarship Council under Grant 202606690023. \textit{(Corresponding author: Yang Li.)}

}
}

%
%

\markboth{}%
{Shell \MakeLowercase{\textit{et al.}}: Bare Demo of IEEEtran.cls for Journals}
%



\maketitle
\begin{abstract}
Insertion-deletion (insdel) codes correct synchronization errors. By the half-Singleton bound, an ${[n,2]}_q$ code has minimum insdel distance at most $2n-4$. The smallest field size of two-dimensional Reed--Solomon (RS) codes attaining this bound is $\Theta(n^3)$. Unlike the minimum Hamming distance, the minimum insdel distance of an RS code depends on the ordering of its evaluation points. An evaluation set is called ordering-free if every ordering of its points gives an RS code attaining the half-Singleton bound. In this paper, we first show that for $n\ge6$ every ${[n,2]}_q$ code attaining the half-Singleton bound is an RS code, and hence $q\ge\binom{n}{3}+2$. Secondly, we show that an ordering-free evaluation set with $n\ge6$ points in $\mathbb{F}_q$ satisfies $n<q^{1/3}+2$, and that it has at most $r+1$ points if $q=r^3$ for a prime power $r\ge3$. Finally, by using cross-ratios, we construct an ordering-free evaluation set with $r+1$ points in $\mathbb{F}_{r^3}$ for every prime power $r$. Hence, the bound $r+1$ is attained for $r\ge3$. Moreover, the constructed set is even robust against permutations combined with insdel errors.
\end{abstract}

%
\IEEEpeerreviewmaketitle

\section{Introduction}\label{section001}
In communication and storage channels prone to synchronization disturbances, a symbol of the transmitted word may be lost, or a spurious symbol may appear between two correct ones. To address these synchronization errors, \emph{insertion-deletion} (\emph{insdel}) codes were proposed in the 1960s by Varshamov and Tenengolts~\cite{R07} and Levenshtein~\cite{LY07,R05}. They were further developed by Tenengolts~\cite{LY311}. Nowadays, insertion and deletion errors arise in various domains, including racetrack memories~\cite{LY290}, language processing~\cite{LY289} and DNA-based data storage~\cite{LY292,LY297}. Many works constructed codes correcting one, two or a few insdel errors with low redundancy~\cite{LY31,LY71,LY131,LY299,LY285}.

For linear insdel codes, Cheng \emph{et al.}~\cite[Theorem~1.2]{LY218} proposed the \emph{half-Singleton bound}. They also obtained the first asymptotically good binary linear insdel codes~\cite[Theorems~1.1 and~1.4]{LY218}. Chen~\cite[Section~II]{LY27} gave the non-asymptotic version of the half-Singleton bound with a simple proof. Ji \emph{et al.}~\cite[Theorem~3.1]{LY30} obtained the strict half-Singleton bound for linear codes that do not contain the all-one codeword. Moreover, the rate-distance tradeoffs of linear insdel codes were improved by Con \emph{et al.}~\cite{LY152}, and the high-noise and high-rate regimes were studied in~\cite{LY103}.

Reed--Solomon (RS) codes have a simple algebraic structure, and they have been used in many scenarios, such as secret sharing~\cite{LY306} and distributed storage~\cite{LY303}. The decoding of RS codes against substitution errors has been studied extensively~\cite{LY309,LY300}. Besides substitution errors, generalized RS codes can also correct deletions, and hence insdel errors. Safavi-Naini and Wang~\cite[Theorem~3]{LY195} gave an algebraic condition for generalized RS codes to correct deletions. Wang \emph{et al.}~\cite[Theorem~3.2]{LY107} obtained ${[5,2]}_q$ generalized RS codes correcting a single deletion, and Tonien and Safavi-Naini~\cite[Corollary~1]{LY106} showed that ${[n,k]}_q$ RS codes with suitable evaluation points correct a number of deletions logarithmic in $n$. Con \emph{et al.}~\cite{LY65} gave a sufficient algebraic condition for RS codes to attain the half-Singleton bound. They proved the existence of ${[n,k]}_q$ RS codes attaining this bound with $q=n^{O(k)}$. These codes are the first linear codes attaining this bound for general $k$~\cite{LY03}. Con \emph{et al.}~\cite{LY104} showed that random RS codes over fields of linear size come close to this bound. For recent developments on insdel codes, in particular on (generalized) RS codes, we refer to~\cite{LY234,LY228,LY65,LY03,LY24,LY49} and the references therein.

Among these works, the two-dimensional case has received particular attention. In this case, one focuses on ${[n,2]}_q$ RS codes with minimum insdel distance $2n-4$, the largest possible value, over fields as small as possible~\cite{LY120,LY66,LY01,LY65,LY03}. The known constructions are listed in Table~\ref{table001}. Con \emph{et al.}~\cite[Proposition~37]{LY65} proved that such codes require $q=\Omega(n^3)$. Con \emph{et al.}~\cite{LY03} gave explicit constructions over cubic extension fields $\mathbb{F}_{r^3}$ of size $\Theta(n^3)$. For decoding, Singhvi~\cite[Theorem~3]{LY124} gave an algorithm that corrects $n-3$ deletions for every ${[n,2]}_q$ RS code attaining the half-Singleton bound. Singhvi~\cite[Theorem~2]{LY124} also gave an algorithm with running time $O(n)$ for the construction in~\cite{LY03}. Besides, Do Duc \emph{et al.}~\cite[Theorem~2]{LY66} and Liu and Tjuawinata~\cite[Theorem~1]{LY193} constructed two-dimensional RS codes whose relative minimum insdel distance is asymptotically at least $1-\varepsilon$. Beelen \emph{et al.}~\cite{LY81} studied full-length two-dimensional RS codes against a linear fraction of insdel errors. Wang and Ding~\cite[Theorem~4.3]{LY64} obtained two-dimensional near MDS codes with minimum insdel distance $2n-6$ from twisted RS codes.

\begin{table}[ht]
\centering
\begin{threeparttable}
\caption{Explicit ${[n,2]}_q$ RS Codes Attaining the Half-Singleton Bound}\label{table001}
\begin{tabular*}{0.8\textwidth}{@{\extracolsep{\fill}}llll}
\toprule
Work & Field size $q$ & Largest length over $\mathbb{F}_{r^3}$\tnote{$\dagger$} & Orderings\\
\midrule
\cite[Theorem~3]{LY66} & $\exp(O(n^2))$ & -- & every ordering\\
\cite[Corollary~C]{LY120} & $\exp(O(2^n))$ & -- & one ordering\\
\cite[Theorem~4.12]{LY01} & $O(n^5)$ & -- & one ordering\\
\cite[Theorem~30]{LY65} & $O(n^4)$ & -- & one ordering\\
\cite[Propositions~2 and~3]{LY03} & $O(n^3)$ & $r-1$ & every ordering\\
Our Theorem~\ref{theorem025} & $O(n^3)$ & $r+1$\tnote{$\ddagger$} & every ordering\\
\bottomrule
\end{tabular*}
\begin{tablenotes}\footnotesize
\item[$\dagger$] The integer $r$ is a prime power.
\item[$\ddagger$] For $r\ge3$, the length $r+1$ attains the upper bound $r+1$ for ordering-free evaluation sets in Theorem~\ref{theorem021}(b).
\end{tablenotes}
\end{threeparttable}
\end{table}
\subsection{Motivations}
Inspired by the works mentioned above, we study two-dimensional linear codes attaining the half-Singleton bound, with emphasis on RS codes and the orderings of their evaluation points. We have three main motivations.  

\textbf{Motivation 1.} Is every two-dimensional linear code attaining the half-Singleton bound an RS code? Let $\mathscr{C}$ be an ${[n,2]}_q$ linear code. Since the minimum insdel distance $d_I(\mathscr{C})$ and the minimum Hamming distance $d_H(\mathscr{C})$ satisfy $d_I(\mathscr{C})\le2d_H(\mathscr{C})$~\cite[Lemma~III.2]{LY120}, attaining the half-Singleton bound only forces $d_H(\mathscr{C})\ge n-2$. Hence, $\mathscr{C}$ does not need to be MDS, and in particular it does not need to be an RS code.

Let $n\ge4$. By the strict half-Singleton bound~\cite[Theorem~3.1]{LY30}, if $\mathscr{C}$ attains the half-Singleton bound, then $\mathscr{C}$ contains the all-one vector $\mathbf{1}_n$. Therefore, $\mathscr{C}=\{a\mathbf{1}_n+b\overline{\boldsymbol{\alpha}}:a,b\in\mathbb{F}_q\}$ for some $\overline{\boldsymbol{\alpha}}\in\mathbb{F}_q^n$. The code $\mathscr{C}$ is an RS code if and only if the entries of $\overline{\boldsymbol{\alpha}}$ are pairwise distinct. 

Moreover, Con \emph{et al.}~\cite[Section~1.4]{LY104} asked for a lower bound on the field size of linear codes attaining the half-Singleton bound. For ${[n,2]}_q$ RS codes attaining this bound, the lower bound $q\ge\binom{n}{3}-1$ is known~\cite[Remark~2]{LY03}. If every ${[n,2]}_q$ linear code attaining the half-Singleton bound is an RS code, then this lower bound also holds for all these linear codes. These observations lead to the following problem.

\begin{problem}\label{problem001}
For sufficiently large $n$, is every ${[n,2]}_q$ linear code attaining the half-Singleton bound an RS code?
\end{problem}    

\textbf{Motivation 2.} An RS code is defined by an evaluation vector, not by an evaluation set. The same points in different orderings can give RS codes with different minimum insdel distances. This phenomenon is not special to RS codes. Chen~\cite[Section~III]{LY27} observed that the coordinate ordering strongly affects the minimum insdel distance. Xie \emph{et al.}~\cite[Corollary~2.8]{LY29} obtained binary codes with small minimum insdel distance by reordering coordinates. Qiu \emph{et al.}~\cite{LY215} stressed that their results on minimum insdel distances depend on the coordinate ordering.

Con~\cite{LY46} studied RS codes against an adversary that permutes a codeword and then applies insertions and deletions. By~\cite[Remark~1]{LY46}, this model falls within the framework of Kova\v{c}evi\'{c} and Tan~\cite{LY295} for permutation channels, where only the multiset of transmitted symbols carries information. By~\cite[Theorem~4]{LY46}, the two-dimensional RS codes of~\cite{LY03} are robust against $n-3$ such errors. We call an evaluation set with this robustness \emph{permutation-robust}. A weaker requirement is that every ordering of the evaluation set gives a two-dimensional RS code attaining the half-Singleton bound. We call such an evaluation set \emph{ordering-free}. The two notions differ in how the order is chosen. For an ordering-free set, the order of the coordinates is arbitrary, but it is fixed in advance as part of the code. For a permutation-robust set, the adversary permutes each transmitted codeword before applying insertions and deletions. Every permutation-robust evaluation set is ordering-free, and the converse fails (Example~\ref{example019}).

For ${[n,2]}_q$ RS codes attaining the half-Singleton bound, the smallest field size is $\Theta(n^3)$~\cite{LY65,LY03}. However, the constant factor is far from being determined. By~\cite[Remark~2]{LY03}, such codes satisfy $q\ge\binom{n}{3}-1$, which is about $n^3/6$. The explicit constructions in~\cite{LY03} use $q=(n+1)^3$. Over $\mathbb{F}_{r^3}$, the lower bound allows lengths up to about $6^{1/3}r\approx1.817\,r$, while the constructions in~\cite{LY03} have length at most $r-1$. In Proposition~\ref{proposition018}, a simple count of ratios shows that a permutation-robust evaluation set in $\mathbb{F}_{r^3}$ has at most $r+1$ points. However, this count does not apply to ordering-free evaluation sets. This leads to the following problem.

\begin{problem}\label{problem002}
How large can an ordering-free evaluation set in $\mathbb{F}_q$ be? In particular, can an ordering-free evaluation set in $\mathbb{F}_{r^3}$ have more than $r+1$ points?
\end{problem}

\textbf{Motivation 3.} All explicit evaluation sets of two-dimensional RS codes in~\cite{LY03} have at most $r-1$ points. They are permutation-robust, but they have two points fewer than the upper bound $r+1$.

Permutation-robust evaluation sets have an application in cryptography. In fully anonymous secret sharing, the secret is reconstructed from shares without the identities of the participants. Using the evaluation sets of~\cite{LY03}, Con~\cite[Corollary~2]{LY46} obtained an explicit fully anonymous $(1,3,n)$ ramp secret sharing scheme over a field of size $O(n^3)$. A permutation-robust evaluation set in $\mathbb{F}_{r^3}$ with more points gives such a scheme with more participants over the same field. This leads to the following problem.

\begin{problem}\label{problem003}
For every prime power $r$, can one construct a permutation-robust evaluation set of two-dimensional RS codes in $\mathbb{F}_{r^3}$ with $r+1$ points?
\end{problem}

\subsection{Our Main Contributions}
In this paper, we study two-dimensional RS codes over $\mathbb{F}_q$ attaining the half-Singleton bound using ratios and cross-ratios of evaluation points. We pay particular attention to the case $q=r^3$, where $r$ is a prime power. In Lemma~\ref{lemma007}, we give a direct criterion in terms of ratios, which is convenient to use. It can be derived from the known criteria in~\cite[Lemma~4.10]{LY01} and~\cite[Proposition~14]{LY65}. Our main contributions are as follows.
\begin{enumerate}
\item Building on the strict half-Singleton bound, we show that for $n\ge6$ every ${[n,2]}_q$ linear code attaining the half-Singleton bound is an RS code (Theorem~\ref{theorem009}), and the condition $n\ge6$ cannot be removed (Example~\ref{example010}). This answers \textbf{Problem}~\ref{problem001}. Hence, for $n\ge6$, the restriction to RS codes is no loss, and the deletion decoding algorithm of~\cite[Theorem~3]{LY124} for RS codes attaining the half-Singleton bound applies to every two-dimensional linear code of length $n\ge6$ attaining this bound. Moreover, every such code needs $q\ge\binom{n}{3}+2$, which answers the case $k=2$ and $n\ge6$ of the question in~\cite[Section~1.4]{LY104}.
\item Permutation-robust sets in $\mathbb{F}_{r^3}$ have at most $r+1$ points by a simple count (Proposition~\ref{proposition018}). These sets are exactly the sets for which the bound of~\cite[Equation~(10)]{LY66} gives $d_I=2n-4$ in every ordering (Lemma~\ref{lemma017}(a)). For every $q$, an ordering-free set in $\mathbb{F}_q$ with $n\ge6$ points satisfies $n<q^{1/3}+2$ (Theorem~\ref{theorem021}(a)). For $r\ge3$, the bound $r+1$ of Proposition~\ref{proposition018} also holds for ordering-free sets in $\mathbb{F}_{r^3}$, i.e., no evaluation set in $\mathbb{F}_{r^3}$ with more than $r+1$ points gives RS codes attaining the half-Singleton bound in every ordering (Theorem~\ref{theorem021}(b)). For $r=2$, there is an ordering-free set in $\mathbb{F}_8$ with $r+2=4$ points (Remark~\ref{remark003}).
\item The image of $\mathbb{P}^1(\mathbb{F}_r)$ under a suitable M\"obius transformation of $\mathbb{P}^1(\mathbb{F}_{r^3})$ is a permutation-robust set with $r+1$ points, and its ratios realize every element of $\mathbb{F}_{r^3}\setminus\mathbb{F}_r$ exactly once (Lemma~\ref{lemma024} and Theorem~\ref{theorem025}). This gives explicit ${[r+1,2]}_{r^3}$ RS codes attaining the half-Singleton bound in every ordering and for every prime power $r$, two longer than the constructions of~\cite{LY03} over the same field, and hence an answer to \textbf{Problem}~\ref{problem003}. This construction attains the bound of Proposition~\ref{proposition018} for every $r$, and that of Theorem~\ref{theorem021}(b) for $r\ge3$.
\item A translate of the permutation-robust set in Theorem~\ref{theorem025} that avoids $0$ gives an explicit fully anonymous $(1,3,r+1)$ ramp secret sharing scheme over $\mathbb{F}_{r^3}$ in the sense of~\cite{LY46} (Corollary~\ref{corollary026}). Over the same field, it has two more participants than the scheme of~\cite[Corollary~2]{LY46}, and by Proposition~\ref{proposition018}, $r+1$ is the largest number of participants of such schemes from two-dimensional RS codes over $\mathbb{F}_{r^3}$ by the construction of~\cite[Corollary~1]{LY46}.
\end{enumerate}
\subsection{Organization of This Paper}
This paper is organized as follows. Section~\ref{section002} reviews insdel distances and known bounds on linear insdel codes. Section~\ref{section003} gives the ratio criterion and shows that two-dimensional linear codes attaining the half-Singleton bound are RS codes for $n\ge6$. Section~\ref{section004} bounds the size of permutation-robust and ordering-free evaluation sets, and Section~\ref{section005} constructs permutation-robust evaluation sets attaining the upper bound $r+1$ and gives an application to secret sharing. Section~\ref{section006} concludes the paper.

\section{Preliminaries}\label{section002}
This section provides the background and known results required for subsequent discussion. 
\subsection{Basic Notation}
\begin{itemize}
\item Let $\mathbb{F}_q$ be the finite field with $q$ elements, where $q$ is a prime power. Let $\mathbb{F}_q^*:=\mathbb{F}_q\setminus\{0\}$, and let $\mathbb{F}_q^n$ be the set of all vectors of length $n$ over $\mathbb{F}_q$.
\item A vector $\mathbf{a}\in \mathbb{F}_q^n$ is written as $\mathbf{a}=\left(a_1, \ldots, a_n\right)$.
\item Let $\mathbf{0}_n$ and $\mathbf{1}_n$ be the all-zero and all-one row vectors of length $n$, respectively. 
\item For a positive integer $n$, let $[n]:=\{1,2,\dots,n\}$.
\item For a finite set $\mathcal{A}$, let $|\mathcal{A}|$ denote the number of elements of $\mathcal{A}$.
\item For a set $\mathcal{A}$, let $\mathcal{T}(\mathcal{A}):=\{(a,b,c)\in\mathcal{A}^3: a,b,c \text{ pairwise distinct}\}$ be the set of ordered triples of distinct elements of $\mathcal{A}$. If $\mathcal{A}$ is finite, then $|\mathcal{T}(\mathcal{A})|=|\mathcal{A}|(|\mathcal{A}|-1)(|\mathcal{A}|-2)$.
\item An \emph{affine map} of $\mathbb{F}_q$ is a map $x\mapsto\lambda x+\mu$ with $\lambda,\mu\in\mathbb{F}_q$ and $\lambda\ne0$. In particular, every affine map is a bijection of $\mathbb{F}_q$.
\end{itemize}
\subsection{Hamming and Insdel Distances of Linear Codes}
The \textit{Hamming distance} $d_H(\mathbf{a},\mathbf{b})$ of $\mathbf{a},\mathbf{b}\in\mathbb{F}_q^n$ is the number of coordinates in which they differ, and the \textit{minimum Hamming distance} of a linear code $\mathscr{C}\subseteq\mathbb{F}_q^n$ is $d_H(\mathscr{C})=\min\{d_H(\mathbf{a},\mathbf{b}):\mathbf{a}\ne\mathbf{b}\in\mathscr{C}\}$.

The {\em insdel distance}, denoted as $d_I(\mathbf{a}, \mathbf{b})$, between two words $\mathbf{a},\mathbf{b}\in \mathbb{F}_q^n$ represents the minimum number of insertions and deletions required to transform $\mathbf{a}$ into $\mathbf{b}$. It satisfies the properties of a distance function, and hence forms a metric.
The \textit{minimum insdel distance} of an insdel code $\mathscr{C} \subseteq \mathbb{F}_q^n$ is defined as 
\[
d_I(\mathscr{C})=\min \{d_I(\mathbf{a}, \mathbf{b}):  \mathbf{a} \neq \mathbf{b} \in \mathscr{C}\}.
\]
An ${[n,k]}_q$ code is a $k$-dimensional linear code in $\mathbb{F}_q^n$, and an ${[n,k,(d_H,d_I)]}_q$ code is an ${[n,k]}_q$ code with minimum Hamming distance $d_H$ and minimum insdel distance $d_I$.
By~\cite[Remark~4]{LY66}, a code $\mathscr{C} \subseteq \mathbb{F}_q^n$ with the minimum insdel distance $d_I$ can correct up to $\left\lfloor\frac{d_I-1}{2}\right\rfloor$ insdel errors.

Following~\cite{LY66,LY132}, we compute insdel distances via common subsequences. A {\em common subsequence} of two vectors $\mathbf{a} = (a_1, \ldots, a_n)$ and $\mathbf{b} = (b_1, \ldots, b_n)\in \mathbb{F}_q^n$ is a sequence of length $l$, where $0 \leq l \leq n$, such that there are increasing indices $1 \leq i_1  < \cdots < i_l \leq n$ and $1 \leq j_1  < \cdots < j_l \leq n$ satisfying $(a_{i_1},  \ldots, a_{i_l}) =  (b_{j_1},  \ldots, b_{j_l})$. A {\em longest common subsequence} $(\mathrm{LCS})$ of $\mathbf{a}$ and $\mathbf{b}$ is a common subsequence that achieves the maximum possible length. The length of a longest common subsequence between $\mathbf{a}$ and $\mathbf{b}$ is denoted by $l_{\mathrm{LCS}}(\mathbf{a}, \mathbf{b})$, i.e., 
$l_{\mathrm{LCS}}(\mathbf{a}, \mathbf{b})=\max \{0 \leq l \leq n : \text{there is a common subsequence of } \mathbf{a}, \mathbf{b}\text{ of length } l\}$. The insdel distance can be computed from $l_{\mathrm{LCS}}$ as follows.

\begin{lemma}\cite[Lemma 1]{LY66}\label{lemma001}
	For $\mathbf{a}, \mathbf{b} \in \mathbb{F}_q^n$, $d_I(\mathbf{a}, \mathbf{b})=2 (n-l_{\mathrm{LCS}}(\mathbf{a}, \mathbf{b}))$.
\end{lemma}    

\subsection{Bounds on Linear Insdel Codes}
In this subsection, we review two bounds on linear insdel codes as follows.
\begin{lemma}[Half-Singleton Bound] \cite[Theorem~1.2]{LY218}, \cite[Section~II]{LY27}\label{lemma002}
	Let $\mathscr{C}$ be an ${[n, k]}_q$ code with $n>k \geq 1$. Then
\begin{equation}\notag
d_I(\mathscr{C}) \leq \max \{2(n-2 k+2), 2\}.
\end{equation}
\end{lemma}
\begin{lemma}[Strict Half-Singleton Bound]\cite[Theorem 3.1]{LY30} \label{lemma003}
	Let $\mathscr{C}$ be an ${[n, k]}_q$ code with $n>k \geq 1$. If $\mathbf{1}_n \notin \mathscr{C}$, then
\begin{equation}\notag
d_I(\mathscr{C}) \leq \max \{2(n-2 k+1), 2\}.
\end{equation}
\end{lemma}

By Lemma~\ref{lemma001}, the equality case of Lemma~\ref{lemma002} can be restated as follows.

\begin{lemma}\label{lemma004}
An ${[n,k]}_q$ code $\mathscr{C}$ with $n\ge 2k-1$ is optimal with respect to the half-Singleton bound if and only if $l_{\mathrm{LCS}}(\mathbf{a},\mathbf{b})\le2k-2$ for all distinct $\mathbf{a},\mathbf{b}\in \mathscr{C}$. 
\end{lemma}    
\section{Optimal Two-Dimensional Linear Insdel Codes}\label{section003}
In this section, we give the ratio criterion for optimal two-dimensional RS codes, and we answer \textbf{Problem}~\ref{problem001}.

An ${[n,k]}_q$ code $\mathscr{C}$ with $d_H(\mathscr{C})=n-k+1$ is called a maximum distance separable (MDS) code under the Hamming metric. RS codes are MDS codes, and they are defined as follows.
\begin{definition}\label{definition005}
Let $\boldsymbol{ \alpha }=\left(\alpha_1,\ldots,\alpha_n\right)$ with pairwise distinct $\alpha_1,  \ldots, \alpha_n\in\mathbb{F}_q$. If $k \leq n \leq q$, then the $k$-dimensional {\rm RS} code associated with the evaluation vector $\boldsymbol{ \alpha }$ is defined as
\[
\mathrm{RS}_{n, k}(\boldsymbol{\alpha}):=  \left\{\left( f\left(\alpha_1\right), \ldots, f\left(\alpha_n\right)\right): f(x) \in \mathbb{F}_q[x], \deg(f(x)) \leq k-1\right\}.
\]
\end{definition}

This paper focuses on two-dimensional RS codes, for which we use the following definition.

\begin{definition}\label{definition006}
Let $n\ge3$, and let $\boldsymbol{\alpha}\in\mathbb{F}_q^n$
be an evaluation vector with pairwise distinct entries.
The evaluation vector $\boldsymbol{\alpha}$ and the code $\mathrm{RS}_{n, 2}(\boldsymbol{\alpha})$
are called optimal if $\mathrm{RS}_{n, 2}(\boldsymbol{\alpha})$ attains the half-Singleton bound,
i.e., $d_I(\mathrm{RS}_{n, 2}(\boldsymbol{\alpha}))=2n-4$.
An evaluation set $\mathcal{S}\subseteq\mathbb{F}_q$ with $|\mathcal{S}|=n\ge3$ is called ordering-free
if every ordering of its points is an optimal evaluation vector.
\end{definition}    

\subsection{The Ratio Criterion for Optimal Two-Dimensional Insdel RS Codes}
Let $\boldsymbol{\alpha}=(\alpha_1,\dots,\alpha_n)\in\mathbb{F}_q^n$ be an evaluation vector with pairwise distinct entries. For distinct $a,b,c\in\mathbb{F}_q$, let $\rho(a,b,c):=(c-a)/(b-a)\in\mathbb{F}_q\setminus\{0,1\}$. A vector $I=(I_1,I_2,I_3)\in[n]^3$ with $I_1<I_2<I_3$ is called an \emph{increasing triple}. For an increasing triple $I$, let $\boldsymbol{\alpha}_I:=(\alpha_{I_1},\alpha_{I_2},\alpha_{I_3})$. Let $\rho_I:=\rho(\alpha_{I_1},\alpha_{I_2},\alpha_{I_3})$, and for $1\le i<j<k\le n$ let $\rho_{ijk}:=\rho(\alpha_i,\alpha_j,\alpha_k)$. For increasing triples $I$ and $J$, let $d_H(I,J)$ denote the number of $t\in\{1,2,3\}$ with $I_t\ne J_t$.

Criteria equivalent to Lemma~\ref{lemma007} are known. Liu and Xing~\cite[Lemma~4.10]{LY01} stated one in terms of affine maps and pairs of increasing triples $I,J$ with $d_H(I,J)\ge2$. Another one can be derived from the case $k=2$ of~\cite[Proposition~14]{LY65} and its ``moreover'' part. Singhvi~\cite[Lemma~2]{LY124} showed that the condition of~\cite[Proposition~14]{LY65} implies that the values $(\alpha_i-\alpha_j)/(\alpha_j-\alpha_k)=1/(\rho_{ijk}-1)$, $i<j<k$, are pairwise distinct. In the ratio form, affine maps and the condition on $d_H(I,J)$ do not appear, and the criterion becomes the injectivity of $I\mapsto\rho_I$ on increasing triples. This is the form used in Corollary~\ref{corollary008} and Section~\ref{section004}. Therefore, we state the criterion in the ratio form and give a direct proof.

\begin{lemma}\label{lemma007}
Let $n\ge3$. The ${[n, 2]}_q$ RS code $\mathscr{C}$ with the evaluation vector $\boldsymbol{\alpha}$ is optimal if and only if the $\binom{n}{3}$ ratios $\rho_{ijk}$, where $1\le i<j<k\le n$, are pairwise distinct.
\end{lemma}
\begin{IEEEproof}
Every codeword of $\mathscr{C}$ has the form $f(\boldsymbol{\alpha})=(f(\alpha_1),\dots,f(\alpha_n))$ with $f(x)=u+vx$, $u,v\in\mathbb{F}_q$. If $v=0$, then $f(\alpha_1)=\cdots=f(\alpha_n)$. If $v\ne0$, then $f(\alpha_1),\dots,f(\alpha_n)$ are pairwise distinct, and for distinct $x,y,z\in\mathbb{F}_q$ we have
\begin{equation}\label{eq001}
\rho(f(x),f(y),f(z))=\frac{v(z-x)}{v(y-x)}=\rho(x,y,z).
\end{equation}
By Lemma~\ref{lemma004}, $\mathscr{C}$ is optimal if and only if no two distinct codewords have a common subsequence of length $3$.

First suppose that the ratios $\rho_{ijk}$ are pairwise distinct. Assume, for contradiction, that distinct codewords $f(\boldsymbol{\alpha})$ and $g(\boldsymbol{\alpha})$ have a common subsequence of length $3$, i.e., $f(\boldsymbol{\alpha}_I)=g(\boldsymbol{\alpha}_J)$ for some increasing triples $I$ and $J$. If $f$ is constant, then $g(\boldsymbol{\alpha}_J)$ has three equal entries, which forces $g$ to be constant, and then $f=g$. Hence, $f$ is nonconstant, and in the same way $g$ is nonconstant. By~\eqref{eq001}, we obtain $\rho_I=\rho(f(\alpha_{I_1}),f(\alpha_{I_2}),f(\alpha_{I_3}))=\rho(g(\alpha_{J_1}),g(\alpha_{J_2}),g(\alpha_{J_3}))=\rho_J$. Since the ratios are pairwise distinct, we have $I=J$, which yields $f(\alpha_{I_1})=g(\alpha_{I_1})$ and $f(\alpha_{I_2})=g(\alpha_{I_2})$. Since $\deg(f)\le 1$ and $\deg(g)\le 1$, it follows from $\alpha_{I_1}\ne\alpha_{I_2}$ that $f=g$, which is a contradiction.

Conversely, suppose that $\mathscr{C}$ is optimal. Assume, for contradiction, that $\rho_I=\rho_J$ for some distinct increasing triples $I$ and $J$. Let $\tilde{v}:=(\alpha_{J_2}-\alpha_{J_1})/(\alpha_{I_2}-\alpha_{I_1})$ and $\tilde{u}:=\alpha_{J_1}-\tilde{v}\alpha_{I_1}$, and let $\tilde{f}(x):=\tilde{u}+\tilde{v}x$. Then $\tilde{v}\ne0$, $\tilde{f}(\alpha_{I_1})=\alpha_{J_1}$ and $\tilde{f}(\alpha_{I_2})=\alpha_{J_2}$. For distinct $x,y,z\in\mathbb{F}_q$, we have $z=x+(y-x)\rho(x,y,z)$. Applying this to $\tilde{f}(\alpha_{I_1}),\tilde{f}(\alpha_{I_2}),\tilde{f}(\alpha_{I_3})$ and using~\eqref{eq001} for $\tilde{f}$, we obtain $\tilde{f}(\alpha_{I_3})=\alpha_{J_1}+(\alpha_{J_2}-\alpha_{J_1})\rho_I=\alpha_{J_1}+(\alpha_{J_2}-\alpha_{J_1})\rho_J=\alpha_{J_3}$. Hence, the codewords $\tilde{f}(\boldsymbol{\alpha})$ and $\boldsymbol{\alpha}$ have the common subsequence $\boldsymbol{\alpha}_J$ of length $3$. Since $I\ne J$, there is $t\in\{1,2,3\}$ with $I_t\ne J_t$. Since the entries of $\boldsymbol{\alpha}$ are pairwise distinct, we obtain $\tilde{f}(\alpha_{I_t})=\alpha_{J_t}\ne\alpha_{I_t}$. Therefore, the two codewords are distinct, which contradicts the optimality of $\mathscr{C}$.
\end{IEEEproof}
\begin{remark}\label{remark001}
The condition in Lemma~\ref{lemma007} depends on the ordering of the entries of $\boldsymbol{\alpha}$. For each $3$-subset $\{a,b,c\}$ of the evaluation set, Lemma~\ref{lemma007} uses only the ratio $\rho(a,b,c)$ with $a,b,c$ listed in their order in $\boldsymbol{\alpha}$, which is one of at most six possible values. A permutation of $\boldsymbol{\alpha}$ may change this ratio, and hence whether the $\binom{n}{3}$ ratios are pairwise distinct. Therefore, the same evaluation set may give an optimal RS code in one ordering but not in another (see Example~\ref{example027}).
\end{remark}

\begin{corollary}\label{corollary008}
Let $n\ge3$. Every optimal ${[n,2]}_q$ RS code satisfies $\binom{n}{3}\le q-2$.
\end{corollary}
\begin{IEEEproof}
By Lemma~\ref{lemma007}, the $\binom{n}{3}$ ratios $\rho_{ijk}$ of the code are pairwise distinct elements of $\mathbb{F}_q\setminus\{0,1\}$, and hence $\binom{n}{3}\le q-2$.
\end{IEEEproof}
Corollary~\ref{corollary008} yields $q\ge\binom{n}{3}+2$ for every optimal ${[n,2]}_q$ RS code, which slightly improves the bound $q\ge\binom{n}{3}-1$ noted in~\cite[Remark~2]{LY03}.    

\subsection{Two-Dimensional Linear Insdel Codes Attaining the Half-Singleton Bound}

Let $\mathscr{C}$ be an arbitrary ${[n,2]}_q$ code. Since $d_I(\mathscr{C})\le2d_H(\mathscr{C})$~\cite[Lemma~III.2]{LY120}, attaining the half-Singleton bound, i.e., $d_I(\mathscr{C})=2n-4$, only forces $d_H(\mathscr{C})\ge n-2$. Hence, $\mathscr{C}$ does not need to be MDS.

For $\overline{\boldsymbol{\alpha}}=(\overline{\alpha}_1,\dots,\overline{\alpha}_n)\in\mathbb{F}_q^n$ whose entries are not all equal, let $\mathscr{E}_2(\overline{\boldsymbol{\alpha}}):=\{a\mathbf{1}_n+b\overline{\boldsymbol{\alpha}}:a,b\in\mathbb{F}_q\}$. If the entries of $\overline{\boldsymbol{\alpha}}$ are pairwise distinct, then $\mathscr{E}_2(\overline{\boldsymbol{\alpha}})=\mathrm{RS}_{n,2}(\overline{\boldsymbol{\alpha}})$ by Definition~\ref{definition005}, and it is MDS. Otherwise, let $c$ be an entry that occurs at least twice in $\overline{\boldsymbol{\alpha}}$. Then $\overline{\boldsymbol{\alpha}}-c\mathbf{1}_n$ is a nonzero codeword of weight at most $n-2$, and hence $\mathscr{E}_2(\overline{\boldsymbol{\alpha}})$ is not MDS. Let $n\ge4$. By Lemma~\ref{lemma003}, an ${[n,2]}_q$ code attaining the half-Singleton bound contains $\mathbf{1}_n$, and hence it equals $\mathscr{E}_2(\overline{\boldsymbol{\alpha}})$ for some $\overline{\boldsymbol{\alpha}}\in\mathbb{F}_q^n$ whose entries are not all equal. Therefore, $\mathscr{E}_2(\overline{\boldsymbol{\alpha}})$ is an RS code if and only if the entries of $\overline{\boldsymbol{\alpha}}$ are pairwise distinct. We show that this holds for $n\ge6$.

\begin{theorem}\label{theorem009}
Let $n\ge 6$, and let $\mathscr{C}$ be an ${[n,2]}_q$ code. If $\mathscr{C}$ attains the half-Singleton bound, then $\mathscr{C}$ is an RS code.
\end{theorem}
\begin{IEEEproof}
By Lemma~\ref{lemma004}, for any two distinct codewords $\mathbf{c}_1,\mathbf{c}_2\in\mathscr{C}$, we have
\begin{equation}\label{eq002}
l_{\mathrm{LCS}}(\mathbf{c}_1,\mathbf{c}_2)\leq 2.
\end{equation}
Since $\max\{2(n-3),2\}<2n-4$, Lemma~\ref{lemma003} yields $\mathbf{1}_n\in\mathscr{C}$.
Hence, $\mathscr{C}=\mathscr{E}_2(\overline{\boldsymbol{\alpha}})$ for some $\overline{\boldsymbol{\alpha}}=(\overline{\alpha}_1,\dots,\overline{\alpha}_n)\in\mathbb{F}_q^n$ whose entries are not all equal.
We now prove that $\overline{\alpha}_1,\dots,\overline{\alpha}_n$ are pairwise distinct.

For any $\xi\in\mathbb{F}_q$, since $\overline{\boldsymbol{\alpha}}-\xi \mathbf{1}_n \neq \mathbf{0}_n$, it follows from \eqref{eq002} that
\begin{equation}\label{eq003}
\left|\left\{t \in[n]: \overline{\alpha}_t=\xi\right\}\right| \leq l_{\mathrm{LCS}}(\mathbf{0}_n, \overline{\boldsymbol{\alpha}}-\xi \mathbf{1}_n) \leq 2 .
\end{equation}
For $1\le i<j\le n$, let $\mathcal{Z}_1(i,j):=\{t\in[n]: t<i\}$, $\mathcal{Z}_2(i,j):=\{t\in[n]: i<t<j\}$ and $\mathcal{Z}_3(i,j):=\{t\in[n]: t>j\}$. Each of them consists of consecutive positions, and $\sum_{w=1}^3|\mathcal{Z}_w(i,j)|=n-2$.

\begin{claim}\label{claim001}
For any pair of positions $1\le i<j\le n$ with $\overline{\alpha}_i=\overline{\alpha}_j$ and every $w\in\{1,2,3\}$, we have
\begin{equation}\label{eqclaim001a}
\bigl|\{\overline{\alpha}_t: t\in\mathcal{Z}_w(i,j)\}\bigr|\le1
\end{equation}
and
\begin{equation}\label{eqclaim001b}
|\mathcal{Z}_w(i,j)|\le2.
\end{equation}
\end{claim}
\textit{Proof of Claim~\ref{claim001}:} Let $\xi=\overline{\alpha}_i=\overline{\alpha}_j$, and fix $w\in\{1,2,3\}$. By~\eqref{eq003}, we have $\overline{\alpha}_t\ne\xi$ for $t\in[n]\setminus\{i,j\}$. Assume, for contradiction, that there exist $e,h\in\mathcal{Z}_w(i,j)$ with $\overline{\alpha}_e\neq\overline{\alpha}_h$. Since $\overline{\alpha}_e\ne\xi$, we may let
$
\lambda=\frac{\overline{\alpha}_h-\xi}{\overline{\alpha}_e-\xi},~\mathbf{u}=\overline{\boldsymbol{\alpha}}-\xi \mathbf{1}_n$, and $\mathbf{v}=\lambda \mathbf{u}$.
Then $\mathbf{u},\mathbf{v}\in\mathscr{C}$. Since $\overline{\alpha}_e\neq\overline{\alpha}_h$, we have $\lambda\ne1$, and hence $v_e\ne u_e$, as $u_e\ne0$. Therefore, $\mathbf{u}\neq\mathbf{v}$. Moreover,
$u_i=v_i=u_j=v_j=0$ and $v_e=\lambda\left(\overline{\alpha}_e-\xi\right)=\overline{\alpha}_h-\xi=u_h$.
Since $e,h\in\mathcal{Z}_w(i,j)$, the positions $i,j,h$ of $\mathbf{u}$ and the positions $i,j,e$ of $\mathbf{v}$ occur in the same relative order. Hence, the entries $u_i=v_i$, $u_j=v_j$ and $u_h=v_e$ form a common subsequence of length $3$, and $l_{\mathrm{LCS}}(\mathbf{u},\mathbf{v})\ge3$, which contradicts~\eqref{eq002}. Therefore, all entries of $\overline{\boldsymbol{\alpha}}$ in $\mathcal{Z}_w(i,j)$ are equal, which proves~\eqref{eqclaim001a}. By~\eqref{eqclaim001a} and~\eqref{eq003}, we obtain~\eqref{eqclaim001b}. This proves Claim~\ref{claim001}.

Now assume, for contradiction, that $\overline{\alpha}_i=\overline{\alpha}_j$ for some $1\le i<j\le n$. Since $\sum_{w=1}^3|\mathcal{Z}_w(i,j)|=n-2\ge4$, \eqref{eqclaim001b} gives some $w\in\{1,2,3\}$ with $|\mathcal{Z}_w(i,j)|=2$, i.e., $\mathcal{Z}_w(i,j)=\{e,e+1\}$ for some $e$. By~\eqref{eqclaim001a}, $\overline{\alpha}_e=\overline{\alpha}_{e+1}$. Since the positions $e$ and $e+1$ are adjacent, we have $\mathcal{Z}_2(e,e+1)=\varnothing$, and hence $|\mathcal{Z}_1(e,e+1)|+|\mathcal{Z}_3(e,e+1)|=n-2\ge4$. By~\eqref{eqclaim001b} for the pair $(e,e+1)$, both terms are at most $2$. Hence, $|\mathcal{Z}_1(e,e+1)|=2$, i.e., $\mathcal{Z}_1(e,e+1)=\{1,2\}$. By~\eqref{eqclaim001a} for the pair $(e,e+1)$, we obtain $\overline{\alpha}_1=\overline{\alpha}_2$. However, $|\mathcal{Z}_3(1,2)|=n-2>2$, which contradicts~\eqref{eqclaim001b} for the pair $(1,2)$.

Hence, the entries of $\overline{\boldsymbol{\alpha}}$ are pairwise distinct, and $\mathscr{C}=\mathrm{RS}_{n,2}(\overline{\boldsymbol{\alpha}})$ by Definition~\ref{definition005}.
\end{IEEEproof}
Theorem~\ref{theorem009} answers \textbf{Problem}~\ref{problem001}: for $n\ge6$, every ${[n,2]}_q$ linear code attaining the half-Singleton bound is an RS code. The following example shows that the condition $n\ge6$ cannot be weakened.

\begin{example}\label{example010}
(a) For $n=3$, the code $\{(a,b,0):a,b\in\mathbb{F}_q\}$ attains the half-Singleton bound.

(b) For $n=4$, the binary code generated by $(1,1,0,0)$ and $(0,0,1,1)$ attains the half-Singleton bound~\cite[Table~II]{LY30}.

(c) For $n=5$, let $\mathbb{F}_8=\mathbb{F}_2[\theta]/(\theta^3+\theta+1)$ and $\overline{\boldsymbol{\alpha}}=(1,0,\theta,0,1+\theta)$. By a Magma~\cite{LY92} computation over all pairs of the~$8^2$ codewords, $\mathscr{E}_2(\overline{\boldsymbol{\alpha}})$ is a ${[5,2,(3,6)]}_8$ code.

None of these codes is MDS, since each of them contains a nonzero codeword of weight at most $n-2$. Hence, none of them is an RS code. Moreover, an optimal ${[5,2]}_q$ RS code requires $q\ge\binom{5}{3}+2=12$ by Corollary~\ref{corollary008}, while the code in (c) is over $\mathbb{F}_8$.
\end{example}
By Theorem~\ref{theorem009} and Corollary~\ref{corollary008}, every ${[n,2]}_q$ code with $n\ge6$ attaining the half-Singleton bound satisfies $q\ge\binom{n}{3}+2$. This answers the case $k=2$ and $n\ge6$ of the question in~\cite[Section~1.4]{LY104}. Together with the explicit constructions in~\cite{LY03}, the bound $q\ge\binom{n}{3}+2$ also shows that the smallest field size of ${[n,2]}_q$ codes attaining the half-Singleton bound is $\Theta(n^3)$, as for RS codes.    

\section{Upper Bounds on Permutation-Robust and Ordering-Free Evaluation Sets}\label{section004}
In this section, we bound the sizes of permutation-robust and ordering-free evaluation sets of optimal two-dimensional insdel RS codes. Together with Theorem~\ref{theorem025}, the bound on ordering-free evaluation sets answers \textbf{Problem}~\ref{problem002}.
\subsection{Orderings and Affine Swaps}
Two $3$-subsets $\mathcal{A},\mathcal{B}\subseteq\mathbb{F}_q$ are \emph{affinely equivalent} if $\varphi(\mathcal{A})=\mathcal{B}$ for some affine map $\varphi$. The bijection $\varphi|_{\mathcal{A}}:\mathcal{A}\to \mathcal{B}$ is then called an \emph{affine equivalence}.

Let $\mathcal{G}$ be the group of the six fractional linear maps $\zeta\mapsto\zeta$, $\zeta\mapsto1/\zeta$, $\zeta\mapsto1-\zeta$, $\zeta\mapsto1/(1-\zeta)$, $\zeta\mapsto(\zeta-1)/\zeta$ and $\zeta\mapsto\zeta/(\zeta-1)$ under composition, regarded as rational functions over $\mathbb{F}_q$. The group $\mathcal{G}$ is generated by $\zeta\mapsto1/\zeta$ and $\zeta\mapsto1-\zeta$~\cite[Theorem~25 and its corollary]{LYBOOK028}. We consider the action of $\mathcal{G}$ on $\mathbb{F}_q\setminus\{0,1\}$. For $\zeta\in\mathbb{F}_q\setminus\{0,1\}$, let
\[
\mathcal{O}(\zeta):=\left\{\zeta,\ \frac1\zeta,\ 1-\zeta,\ \frac1{1-\zeta},\ \frac{\zeta}{\zeta-1},\ \frac{\zeta-1}{\zeta}\right\}
\]
be the $\mathcal{G}$-orbit of $\zeta$, and let $\mathcal{P}_q$ be the set of $\mathcal{G}$-orbits. Then $\mathcal{P}_q$ is a partition of $\mathbb{F}_q\setminus\{0,1\}$~\cite[Theorem~1.4A(i)]{LYBOOK029}.
\begin{lemma}\label{lemma011}
Keep the notation above. Then the following statements hold.
\begin{enumerate}[(a)]
\item For distinct $a,b,c\in\mathbb{F}_q$,
\[
\{\rho(a',b',c'):(a',b',c')\text{ is an ordering of }\{a,b,c\}\}=\mathcal{O}(\rho(a,b,c)).
\]
In particular, for a $3$-subset $\mathcal{A}\subseteq\mathbb{F}_q$, the set $\mathcal{O}(\mathcal{A}):=\mathcal{O}(\rho(a,b,c))$ does not depend on the ordering $(a,b,c)$ of $\mathcal{A}$.
\item Let $(a,b,c),(a',b',c')\in\mathcal{T}(\mathbb{F}_q)$. There is a unique affine map $\varphi$ with $(\varphi(a),\varphi(b),\varphi(c))=(a',b',c')$ if and only if $\rho(a,b,c)=\rho(a',b',c')$.
\item Two $3$-subsets $\mathcal{A},\mathcal{B}\subseteq\mathbb{F}_q$ are affinely equivalent if and only if $\mathcal{O}(\mathcal{A})=\mathcal{O}(\mathcal{B})$.
\item $|\mathcal{P}_q|\le\frac{q+8}{6}$.
\end{enumerate}
\end{lemma}
\begin{IEEEproof}
Part (a) follows from $\rho(a,c,b)=1/\rho(a,b,c)$ and $\rho(b,a,c)=1-\rho(a,b,c)$, since $\zeta\mapsto1/\zeta$ and $\zeta\mapsto1-\zeta$ generate $\mathcal{G}$. For (b), affine maps preserve ratios by~\eqref{eq001}, and the unique affine map with $\varphi(a)=a'$ and $\varphi(b)=b'$ satisfies $\varphi(c)=a'+(b'-a')\rho(a,b,c)$. 

For (c), let $(a,b,c)$ be an ordering of $\mathcal{A}$. If $\varphi(\mathcal{A})=\mathcal{B}$ for an affine map $\varphi$, then $(\varphi(a),\varphi(b),\varphi(c))$ is an ordering of $\mathcal{B}$ with ratio $\rho(a,b,c)$ by (b), and hence $\mathcal{O}(\mathcal{A})=\mathcal{O}(\mathcal{B})$ by (a). Conversely, if $\mathcal{O}(\mathcal{A})=\mathcal{O}(\mathcal{B})$, then by (a) some ordering $(a',b',c')$ of $\mathcal{B}$ satisfies $\rho(a',b',c')=\rho(a,b,c)$. By (b), there is an affine map $\varphi$ with $(\varphi(a),\varphi(b),\varphi(c))=(a',b',c')$, and hence $\varphi(\mathcal{A})=\mathcal{B}$.

For (d), the Cauchy--Frobenius lemma~\cite[Theorem~1.7A]{LYBOOK029} gives
\[
6|\mathcal{P}_q|=\sum_{g\in\mathcal{G}}\bigl|\{\zeta\in\mathbb{F}_q\setminus\{0,1\}: g(\zeta)=\zeta\}\bigr|.
\]
The identity map $\zeta\mapsto\zeta$ fixes all $q-2$ points. Let $g(\zeta)=(\beta_1\zeta+\beta_2)/(\beta_3\zeta+\beta_4)$ be another element of $\mathcal{G}$. The equation $g(\zeta)=\zeta$ gives $\beta_3\zeta^2+(\beta_4-\beta_1)\zeta-\beta_2=0$. This polynomial is nonzero, and hence $g$ has at most two fixed points. Therefore, $6|\mathcal{P}_q|\le q-2+5\cdot2=q+8$.
\end{IEEEproof}

\begin{definition}\label{definition012}
Let $\boldsymbol{\alpha}$ be an ordering of a finite set $\mathcal{S}$, i.e., a vector whose entries are the elements of $\mathcal{S}$, each appearing once. For $3$-subsets $\mathcal{A},\mathcal{B}$ of $\mathcal{S}$, let $I$ and $J$ be the increasing triples with $\mathcal{A}=\{\alpha_{I_1},\alpha_{I_2},\alpha_{I_3}\}$ and $\mathcal{B}=\{\alpha_{J_1},\alpha_{J_2},\alpha_{J_3}\}$. A bijection $\varphi:\mathcal{A}\to\mathcal{B}$ is \emph{order-preserving under $\boldsymbol{\alpha}$} if $\varphi(\boldsymbol{\alpha}_I)=\boldsymbol{\alpha}_J$.
\end{definition}
\begin{corollary}\label{corollary013}
Let $\mathcal{S}\subseteq\mathbb{F}_q$ with $|\mathcal{S}|\ge3$. An ordering $\boldsymbol{\alpha}$ of $\mathcal{S}$ is optimal if and only if no affine equivalence between distinct $3$-subsets of $\mathcal{S}$ is order-preserving under $\boldsymbol{\alpha}$. In particular, $\mathcal{S}$ is ordering-free if and only if no affine equivalence between distinct $3$-subsets of $\mathcal{S}$ is order-preserving under any ordering of $\mathcal{S}$.
\end{corollary}
\begin{IEEEproof}
Let $\boldsymbol{\alpha}$ be an ordering of $\mathcal{S}$. The map $I\mapsto\{\alpha_{I_1},\alpha_{I_2},\alpha_{I_3}\}$ is a bijection from the set of increasing triples onto the set of $3$-subsets of $\mathcal{S}$. Let $\mathcal{A}$ and $\mathcal{B}$ correspond to $I$ and $J$, respectively. Then $I\ne J$ if and only if $\mathcal{A}\ne\mathcal{B}$.

By Lemma~\ref{lemma007}, $\boldsymbol{\alpha}$ is not optimal if and only if $\rho_I=\rho_J$ for some distinct increasing triples $I$ and $J$. By Lemma~\ref{lemma011}(b), this holds if and only if $\varphi(\boldsymbol{\alpha}_I)=\boldsymbol{\alpha}_J$ for some affine map $\varphi$ and some distinct increasing triples $I$ and $J$. By the definition of an affine equivalence and Definition~\ref{definition012}, this holds if and only if some affine equivalence between distinct $3$-subsets of $\mathcal{S}$ is order-preserving under $\boldsymbol{\alpha}$. This proves the first statement.

By Definition~\ref{definition006}, $\mathcal{S}$ is ordering-free if and only if every ordering of $\mathcal{S}$ is optimal. Hence, the second statement follows.
\end{IEEEproof}
The next lemma is purely combinatorial.

\begin{lemma}\label{lemma014}
Let $\mathcal{A}\ne\mathcal{B}$ be $3$-subsets of a finite set $\mathcal{U}$, and let $\varphi:\mathcal{A}\to\mathcal{B}$ be a bijection. Then $\varphi$ is not order-preserving under any ordering of $\mathcal{U}$ if and only if $\varphi$ exchanges two points of $\mathcal{A}\cap\mathcal{B}$.
\end{lemma}
\begin{IEEEproof}
If $\varphi(x)=y$ and $\varphi(y)=x$ with $x,y\in\mathcal{A}\cap\mathcal{B}$, then for every ordering $\boldsymbol{\alpha}$ of $\mathcal{U}$, the points $x$ and $y$ appear in the same relative order in $\boldsymbol{\alpha}_I$ and $\boldsymbol{\alpha}_J$, while $\varphi$ reverses this order. Hence, $\varphi(\boldsymbol{\alpha}_I)\ne\boldsymbol{\alpha}_J$, and $\varphi$ is not order-preserving under any ordering of $\mathcal{U}$. Conversely, suppose that $\varphi$ does not exchange any two points of $\mathcal{A}\cap\mathcal{B}$. It suffices to exhibit an ordering of $\mathcal{A}\cup\mathcal{B}$ under which $\varphi$ is order-preserving, and to append the elements of $\mathcal{U}\setminus(\mathcal{A}\cup\mathcal{B})$. We consider three cases.
\begin{itemize}
    \item \textbf{Case 1: $|\mathcal{A}\cap \mathcal{B}|=0$.} Let $\mathcal{A}=\{a_1,a_2,a_3\}$, and take the ordering $(a_1,a_2,a_3,\varphi(a_1),\varphi(a_2),\varphi(a_3))$.
    \item \textbf{Case 2: $|\mathcal{A}\cap \mathcal{B}|=1$.} Let $\mathcal{A}\cap\mathcal{B}=\{x\}$. If $\varphi(x)=x$, let $\mathcal{A}=\{x,a_1,a_2\}$, and take the ordering $(x,a_1,a_2,\varphi(a_1),\varphi(a_2))$. If $\varphi(x)\ne x$, then $x=\varphi(a_1)$ for a unique $a_1\in\mathcal{A}\setminus\{x\}$, since $x\in\mathcal{B}$. Let $\mathcal{A}=\{a_1,x,a_2\}$, and take the ordering $(a_1,x,a_2,\varphi(x),\varphi(a_2))$.
    \item \textbf{Case 3: $|\mathcal{A}\cap \mathcal{B}|=2$.} Let $\mathcal{A}=\{x,y,a\}$ and $\mathcal{B}=\{x,y,b\}$. There are six bijections from $\mathcal{A}$ to $\mathcal{B}$, and one of them exchanges $x$ and $y$. Since $x$ and $y$ play symmetric roles, each of the other five bijections becomes one of the following three after possibly interchanging the names $x$ and $y$. They are listed below with a suitable ordering of $\mathcal{A}\cup\mathcal{B}$:
    \[
    \begin{array}{l@{\qquad}l}
    x\mapsto x,\ y\mapsto y,\ a\mapsto b: & (x,y,a,b),\\
    x\mapsto x,\ y\mapsto b,\ a\mapsto y: & (x,a,y,b),\\
    x\mapsto y,\ y\mapsto b,\ a\mapsto x: & (a,x,y,b).
    \end{array}
    \]
\end{itemize}
In each case, let $\boldsymbol{\alpha}$ be the displayed ordering followed by the elements of $\mathcal{U}\setminus(\mathcal{A}\cup\mathcal{B})$, and let $I$ and $J$ be the increasing triples with $\mathcal{A}=\{\alpha_{I_1},\alpha_{I_2},\alpha_{I_3}\}$ and $\mathcal{B}=\{\alpha_{J_1},\alpha_{J_2},\alpha_{J_3}\}$. A direct check shows that $\varphi(\boldsymbol{\alpha}_I)=\boldsymbol{\alpha}_J$. Therefore, $\varphi$ is order-preserving under $\boldsymbol{\alpha}$ by Definition~\ref{definition012}.
\end{IEEEproof}

\begin{lemma}\label{lemma015}
Let $\mathcal{S}\subseteq\mathbb{F}_q$, and let $\varphi:\mathcal{A}\to\mathcal{B}$ be an affine equivalence between distinct $3$-subsets $\mathcal{A},\mathcal{B}$ of $\mathcal{S}$. Then $\varphi$ is not order-preserving under any ordering of $\mathcal{S}$ if and only if $\mathcal{A}=\{x,y,w\}$ with $\varphi(x)=y$, $\varphi(y)=x$ and $\varphi(w)=x+y-w$.
\end{lemma}
\begin{IEEEproof}
By Lemma~\ref{lemma014} with $\mathcal{U}=\mathcal{S}$, the condition is that $\varphi$ exchanges two points $x,y\in\mathcal{A}\cap\mathcal{B}$. An affine map with $x\mapsto y$ and $y\mapsto x$ is $t\mapsto x+y-t$. Hence, the third point $w$ of $\mathcal{A}$ satisfies $\varphi(w)=x+y-w$.
\end{IEEEproof}
We call the affine equivalences in Lemma~\ref{lemma015} \textit{affine swaps}.

\subsection{A Counting Bound on Permutation-Robust Evaluation Sets}\label{subsection004002}
\begin{definition}\label{definition016}
A set $\mathcal{S}\subseteq \mathbb{F}_q$ with $|\mathcal{S}|\ge 3$ is \emph{permutation-robust} if the ratios $\rho(a,b,c)$, where $(a,b,c)\in\mathcal{T}(\mathcal{S})$, are pairwise distinct.
\end{definition}
\begin{lemma}\label{lemma017}
Let $\mathcal{S}\subseteq\mathbb{F}_q$ with $|\mathcal{S}|\ge3$.
\begin{enumerate}[(a)]
\item $\mathcal{S}$ is permutation-robust if and only if $|\mathcal{S}\cap\varphi(\mathcal{S})|\le2$ for every affine map $\varphi\ne\mathrm{id}$.
\item If $\mathcal{S}$ is permutation-robust, then $\mathcal{S}$ is ordering-free.
\end{enumerate}
\end{lemma}
\begin{IEEEproof}
(a) First suppose that $\mathcal{S}$ is permutation-robust, and let $\varphi\ne\mathrm{id}$ be an affine map. Assume, for contradiction, that $\mathcal{S}\cap\varphi(\mathcal{S})$ contains three distinct points $a',b',c'$. Then $a=\varphi^{-1}(a')$, $b=\varphi^{-1}(b')$ and $c=\varphi^{-1}(c')$ are distinct points of $\mathcal{S}$. Hence, $(a,b,c),(a',b',c')\in\mathcal{T}(\mathcal{S})$, and $\rho(a,b,c)=\rho(a',b',c')$ by Lemma~\ref{lemma011}(b). Since an affine map $\varphi\ne\mathrm{id}$ fixes at most one point, we have $(a,b,c)\ne(a',b',c')$. This contradicts the assumption that $\mathcal{S}$ is permutation-robust.
 
Conversely, suppose that $|\mathcal{S}\cap\varphi(\mathcal{S})|\le2$ for every affine map $\varphi\ne\mathrm{id}$. Let $(a,b,c),(a',b',c')\in\mathcal{T}(\mathcal{S})$ with $\rho(a,b,c)=\rho(a',b',c')$. By Lemma~\ref{lemma011}(b), there is an affine map $\varphi$ with $(\varphi(a),\varphi(b),\varphi(c))=(a',b',c')$. Then $a',b',c'\in\mathcal{S}\cap\varphi(\mathcal{S})$, and hence $\varphi=\mathrm{id}$. Therefore, $(a,b,c)=(a',b',c')$, and $\mathcal{S}$ is permutation-robust.

(b) Let $\boldsymbol{\alpha}=(\alpha_1,\dots,\alpha_n)$, $n=|\mathcal{S}|$, be an ordering of $\mathcal{S}$. The triples $(\alpha_i,\alpha_j,\alpha_k)$, where $1\le i<j<k\le n$, are distinct elements of $\mathcal{T}(\mathcal{S})$. Since $\mathcal{S}$ is permutation-robust, the ratios $\rho_{ijk}$ are pairwise distinct, and hence $\boldsymbol{\alpha}$ is optimal by Lemma~\ref{lemma007}. Since $\boldsymbol{\alpha}$ is arbitrary, $\mathcal{S}$ is ordering-free.
\end{IEEEproof}
\begin{remark}\label{remark002}
Lemma~\ref{lemma017}(a) relates permutation robustness to the quantity $|\lambda\mathcal{S}\cap(\mathcal{S}+\mu)|$ in the bound of Do Duc~\emph{et al.}~\cite[Equation~(10)]{LY66}. For the affine map $\varphi(x)=\lambda x-\mu$, we have $|\mathcal{S}\cap\varphi(\mathcal{S})|=|\lambda\mathcal{S}\cap(\mathcal{S}+\mu)|$. The proof of~\cite[Lemma~9]{LY66} shows that the sets of~\cite[Theorem~3]{LY66} satisfy $|\lambda\mathcal{S}\cap(\mathcal{S}+\mu)|\le2$ for all $\lambda,\mu\in\mathbb{F}_q$ with $(\lambda,\mu)\ne(1,0)$. Hence, these sets are permutation-robust by Lemma~\ref{lemma017}(a), and they are ordering-free by Lemma~\ref{lemma017}(b). Moreover, as shown in the proof of Corollary~\ref{corollary026}, permutation-robust sets are exactly the evaluation sets of two-dimensional RS codes that are robust against the $(n-3)$-permutation-insdel adversary of Con~\cite{LY46}, where $n=|\mathcal{S}|$. 
\end{remark}
Counting ordered triples gives an upper bound on the size of permutation-robust sets.
\begin{proposition}\label{proposition018}
A permutation-robust set $\mathcal{S}$ satisfies $|\mathcal{S}|(|\mathcal{S}|-1)(|\mathcal{S}|-2)\le q-2$. In particular, $|\mathcal{S}|\le r+1$ when $q=r^3$ for a prime power $r$.
\end{proposition}
\begin{IEEEproof}
By Definition~\ref{definition016}, the ratios $\rho(a,b,c)$, where $(a,b,c)\in\mathcal{T}(\mathcal{S})$, are pairwise distinct elements of $\mathbb{F}_q\setminus\{0,1\}$. Hence, $|\mathcal{S}|(|\mathcal{S}|-1)(|\mathcal{S}|-2)=|\mathcal{T}(\mathcal{S})|\le q-2$. The number $|\mathcal{T}(\mathcal{S})|$ is increasing in $|\mathcal{S}|$, and for $|\mathcal{S}|=r+2$ it equals $r^3+3r^2+2r>r^3-2$. Therefore, $|\mathcal{S}|\le r+1$ when $q=r^3$.
\end{IEEEproof}    

\subsection{An Upper Bound on Ordering-Free Evaluation Sets}\label{subsection004003}
Proposition~\ref{proposition018} does not apply to ordering-free sets. Indeed, the converse of Lemma~\ref{lemma017}(b) fails, and an ordering-free set in $\mathbb{F}_{r^3}$ can have more than $r+1$ points, as the following example shows.
\begin{example}\label{example019}
Let $\mathbb{F}_8=\mathbb{F}_2[\theta]/(\theta^3+\theta+1)$ and $\mathcal{S}=\{0,1,\theta,1+\theta\}$. By a Magma~\cite{LY92} computation, each of the $24$ orderings of $\mathcal{S}$ gives an optimal ${[4,2]}_8$ RS code, i.e., $\mathcal{S}$ is ordering-free. However, $\rho(0,1,\theta)=\rho(1,0,1+\theta)=\theta$, and hence $\mathcal{S}$ is not permutation-robust.
\end{example}
In Theorem~\ref{theorem021}, we show that an ordering-free set in $\mathbb{F}_q$ with $n\ge6$ points satisfies $n<q^{1/3}+2$, and that ordering-free sets in $\mathbb{F}_{r^3}$ still have at most $r+1$ points for $r\ge3$. For this, Lemma~\ref{lemma020} first bounds the number of affine equivalence classes of their $3$-subsets.

For a set $\mathcal{S}\subseteq\mathbb{F}_q$, let
\[
\mathcal{M}_3(\mathcal{S})
:=
\bigl\{
[\mathcal{A}]_{\sim_{\mathrm{aff}}}: \mathcal{A}\subseteq \mathcal{S},\ |\mathcal{A}|=3
\bigr\},
\]
where $[\mathcal{A}]_{\sim_{\mathrm{aff}}}$ is the affine equivalence class
of $\mathcal{A}$ among the $3$-subsets of $\mathcal{S}$.

\begin{lemma}\label{lemma020}
Let $\mathcal{S}\subseteq\mathbb{F}_q$ be ordering-free, with $|\mathcal{S}|=n\ge3$.
Then
\[
|\mathcal{M}_3(\mathcal{S})|
\ge \binom{n}{3}-\frac{n(n-1)}4.
\]
Moreover, if $q$ is odd, then $\binom{n}{3}-|\mathcal{M}_3(\mathcal{S})|$ is even.
\end{lemma}

\begin{IEEEproof}
Since $\mathcal{S}$ is ordering-free, Corollary~\ref{corollary013} and Lemma~\ref{lemma015} show that every affine equivalence between distinct $3$-subsets of $\mathcal{S}$ is an affine swap. An affine swap that exchanges two points with sum $s$ is induced by the affine map
\begin{equation}\label{eq004}
\tau_s(u)=s-u .
\end{equation}
We distinguish two cases according to whether $q$ is odd or even.

\textbf{Case 1:} $q$ is odd. Every $4$-subset $\mathcal{H}$ of $\mathcal{S}$ has at most one partition into two pairs with equal sums. Indeed, let $\sigma$ be the sum of the elements of $\mathcal{H}$, and let $\{\{x,y\},\{z,w\}\}$ be such a partition. Then $2(x+y)=\sigma$. Since $q$ is odd, we obtain $y=\sigma/2-x$. Hence, in such a partition, the partner of each $x\in\mathcal{H}$ is $\sigma/2-x$, and the partition is unique. Let $N_1$ be the number of $4$-subsets of $\mathcal{S}$ that have such a partition.

\begin{claim}\label{claim002}
Every affine equivalence class contains at most two $3$-subsets.
\end{claim}

\textit{Proof of Claim~\ref{claim002}:} Assume, for contradiction, that an affine equivalence class contains three distinct $3$-subsets $\mathcal{A},\mathcal{B},\mathcal{B}'$. By the first sentence of this proof and~\eqref{eq004}, there are $s,t\in\mathbb{F}_q$ with $\mathcal{B}=\tau_s(\mathcal{A})$ and $\mathcal{B}'=\tau_t(\mathcal{A})$. Since $\mathcal{B}\ne\mathcal{B}'$, we have $s\ne t$. Since $\tau_s(\tau_s(u))=u$ for all $u\in\mathbb{F}_q$, we have $\tau_s(\mathcal{B})=\mathcal{A}$. Hence, the map $u\mapsto\tau_t(\tau_s(u))$ carries $\mathcal{B}$ onto $\mathcal{B}'$, and
\[
\tau_t(\tau_s(u))=u+(t-s).
\]
By Lemma~\ref{lemma015}, this map exchanges two points $x,y\in\mathcal{B}\cap\mathcal{B}'$. Then $x+(t-s)=y$ and $y+(t-s)=x$, which yields $2(t-s)=0$. This is impossible, since $q$ is odd and $s\ne t$. This completes the proof of Claim~\ref{claim002}.

We now count the set $\mathcal{K}$ of affine equivalence classes containing exactly two $3$-subsets. Let
\[\mathcal{Q}:=\Bigl\{(\mathcal{H},\mathcal{L}):\ \mathcal{L}\subseteq\mathcal{H}\subseteq\mathcal{S},\ |\mathcal{L}|=2,\ |\mathcal{H}|=4,\ \sum_{u\in\mathcal{L}}u=\sum_{u\in\mathcal{H}\setminus\mathcal{L}}u\Bigr\}.
\]
If $(\mathcal{H},\mathcal{L})\in\mathcal{Q}$, then $\{\mathcal{L},\mathcal{H}\setminus\mathcal{L}\}$ is a partition of $\mathcal{H}$ into two pairs with equal sums, and $(\mathcal{H},\mathcal{H}\setminus\mathcal{L})\in\mathcal{Q}$. Since such a partition of $\mathcal{H}$ is unique, each $4$-subset counted by $N_1$ occurs in exactly two elements of $\mathcal{Q}$. Hence, $|\mathcal{Q}|=2N_1$.

Let $(\mathcal{H},\mathcal{L})\in\mathcal{Q}$, and let $\mathcal{L}=\{x,y\}$ and $\mathcal{H}\setminus\mathcal{L}=\{z,w\}$. Then $x+y=z+w$. By~\eqref{eq004}, we have $\tau_{x+y}(x)=y$, $\tau_{x+y}(y)=x$ and $\tau_{x+y}(z)=w$. Hence, $\tau_{x+y}$ carries $\{x,y,z\}$ onto $\{x,y,w\}$. By Claim~\ref{claim002}, the affine equivalence class of $\{x,y,z\}$ is $\{\{x,y,z\},\{x,y,w\}\}$, which consists of the two $3$-subsets $\mathcal{L}\cup\{z\}$ and $\mathcal{L}\cup\{w\}$ of $\mathcal{H}$ containing $\mathcal{L}$. This defines a map
\begin{equation}\label{eq005}
\mathcal{Q}\to\mathcal{K},~ (\mathcal{H},\mathcal{L})\mapsto\{\mathcal{L}\cup\{z\},\mathcal{L}\cup\{w\}\}.
\end{equation}
Since $z\ne w$, we have
$(\mathcal{L}\cup\{z\})\cup(\mathcal{L}\cup\{w\})=\mathcal{H}
\text{ and }
(\mathcal{L}\cup\{z\})\cap(\mathcal{L}\cup\{w\})=\mathcal{L}$.
Thus $(\mathcal{H},\mathcal{L})$ is determined by its image under~\eqref{eq005}. Therefore, the map~\eqref{eq005} is injective.

We show that the map~\eqref{eq005} is surjective. Let $\{\mathcal{A},\mathcal{B}\}\in\mathcal{K}$. By Lemma~\ref{lemma015}, we have $\mathcal{A}=\{x,y,z\}$ and $\mathcal{B}=\{x,y,w\}$ with $x+y=z+w$. Since $\mathcal{A}\ne\mathcal{B}$, we obtain $(\mathcal{A}\cup\mathcal{B},\mathcal{A}\cap\mathcal{B})\in\mathcal{Q}$, and its image under~\eqref{eq005} is $\{\mathcal{A},\mathcal{B}\}$.

Therefore, the map~\eqref{eq005} is a bijection, and we obtain $|\mathcal{K}|=|\mathcal{Q}|=2N_1$. By Claim~\ref{claim002}, every affine equivalence class contains one or two $3$-subsets. Hence, $\binom{n}{3}=|\mathcal{M}_3(\mathcal{S})|+|\mathcal{K}|$, and we obtain
\begin{equation}\label{eq006}
|\mathcal{M}_3(\mathcal{S})|=\binom{n}{3}-2N_1 .
\end{equation}
In particular, $\binom{n}{3}-|\mathcal{M}_3(\mathcal{S})|$ is even.

It remains to bound $N_1$. For each $d\in\mathbb{F}_q^*$, let $\mathcal{D}_d:=\{a\in \mathcal{S}:a+d\in \mathcal{S}\}$. Each ordered pair of distinct points of $\mathcal{S}$ determines exactly one difference $d$. Hence
\begin{equation}\label{eq007}
\sum_{d\ne0}|\mathcal{D}_d|=n(n-1).
\end{equation}
Let $\mathcal{W}_d$ be the set of unordered pairs $\{u,v\}\subseteq \mathcal{D}_d$ for which $u,u+d,v,v+d$ are pairwise distinct, and let $h_d:=|\mathcal{W}_d|$. Let $\Omega$ be the set of pairs $(d,\{u,v\})$ with $d\in\mathbb{F}_q^*$ and $\{u,v\}\in\mathcal{W}_d$. Then $|\Omega|=\sum_{d\ne0}h_d$. For $(d,\{u,v\})\in\Omega$, the elements $u,u+d,v,v+d$ are pairwise distinct and lie in $\mathcal{S}$, by the definitions of $\mathcal{W}_d$ and $\mathcal{D}_d$. Moreover, $\{u,u+d,v,v+d\}$ does not depend on the order of $u$ and $v$. Hence, the map
\begin{equation}\label{eq008}
\Omega\to\{\mathcal{H}\subseteq\mathcal{S}:|\mathcal{H}|=4\},~ (d,\{u,v\})\mapsto\{u,u+d,v,v+d\}
\end{equation}
is well defined. Let $\mathcal{H}=\{x,y,z,w\}$ be counted by $N_1$, with $x+y=z+w$. Since $z-x=y-w$ and $w-x=y-z$, the four elements
\[
(z-x,\{x,w\}),\quad (x-z,\{z,y\}),\quad (w-x,\{x,z\}),\quad (x-w,\{w,y\})
\]
of $\Omega$ are mapped to $\mathcal{H}$ by the map~\eqref{eq008}. Since $q$ is odd, $z\ne w$ and $z+w=x+y\ne2x$, the first coordinates $z-x$, $x-z$, $w-x$ and $x-w$ are pairwise distinct. Hence, these four elements of $\Omega$ are distinct. Therefore, each $4$-subset counted by $N_1$ has at least four preimages under the map~\eqref{eq008}, and we obtain
\begin{equation}\label{eq009}
4N_1\le\sum_{d\ne0}h_d .
\end{equation}

\begin{claim}\label{claim003}
For every $d\in\mathbb{F}_q^*$, we have $h_d\le|\mathcal{D}_d|/2$.
\end{claim}

\textit{Proof of Claim~\ref{claim003}:} If $h_d=0$, the claim is trivial. Let $\{u,v\}\in\mathcal{W}_d$. Assume, for contradiction, that there is a point $c\in\mathcal{D}_d\setminus\{u,v\}$, and let $\mathcal{R}=\{u,v,c\}$. Then $\mathcal{R}+d\subseteq\mathcal{S}$. Since $u,u+d,v,v+d$ are pairwise distinct, at most one of $u+d$ and $v+d$ equals $c$. Therefore, $\mathcal{R}+d$ contains a point outside $\mathcal{R}$, and $\mathcal{R}+d\ne\mathcal{R}$. By Corollary~\ref{corollary013} and Lemma~\ref{lemma015}, the translation by $d$ exchanges two points $e,f$ of $\mathcal{R}\cap(\mathcal{R}+d)$. Then $e+d=f$ and $f+d=e$, which yields $2d=0$, and this is impossible. Hence, $\mathcal{D}_d=\{u,v\}$, which yields $|\mathcal{D}_d|=2$ and $h_d=1$. Therefore, $h_d\le|\mathcal{D}_d|/2$, and Claim~\ref{claim003} holds.

Combining~\eqref{eq007}, \eqref{eq009} and Claim~\ref{claim003}, we obtain
\[
4N_1\le\sum_{d\ne0}h_d\le\frac12\sum_{d\ne0}|\mathcal{D}_d|=\frac{n(n-1)}2 .
\]
By~\eqref{eq006}, it follows that $|\mathcal{M}_3(\mathcal{S})|\ge\binom{n}{3}-\frac{n(n-1)}4$.

\textbf{Case 2:} $q$ is even. Let $N_2$ count the $4$-subsets of $\mathcal{S}$ whose elements sum to zero. For a $3$-subset $\mathcal{A}=\{x,y,z\}$ of $\mathcal{S}$, let $s_{\mathcal{A}}:=x+y+z$. Then $s_{\mathcal{A}}\notin\mathcal{A}$, since $x,y,z$ are pairwise distinct. By Lemma~\ref{lemma015} and~\eqref{eq004}, every $3$-subset in $[\mathcal{A}]_{\sim_{\mathrm{aff}}}$ other than $\mathcal{A}$ has the form $(\mathcal{A}\setminus\{c\})\cup\{s_{\mathcal{A}}\}$ with $c\in\mathcal{A}$. Hence,
\[
[\mathcal{A}]_{\sim_{\mathrm{aff}}}\subseteq\bigl\{\mathcal{B}\subseteq\mathcal{H}_{\mathcal{A}}:|\mathcal{B}|=3\bigr\},~\text{where }\mathcal{H}_{\mathcal{A}}:=\mathcal{A}\cup\{s_{\mathcal{A}}\}.
\]
Let $\mathcal{K}'$ be the set of affine equivalence classes containing more than one $3$-subset. Let $[\mathcal{A}]_{\sim_{\mathrm{aff}}}\in\mathcal{K}'$, and let $\mathcal{B}\ne\mathcal{A}$ be a $3$-subset in it. Then $s_{\mathcal{A}}\in\mathcal{B}\subseteq\mathcal{S}$ by the inclusion above. Hence, $\mathcal{H}_{\mathcal{A}}$ is a $4$-subset of $\mathcal{S}$, and its elements sum to $2s_{\mathcal{A}}=0$. Thus, $\mathcal{H}_{\mathcal{A}}$ is counted by $N_2$. Moreover, $\mathcal{H}_{\mathcal{A}}=\mathcal{A}\cup\mathcal{B}$ is the union of the $3$-subsets in $[\mathcal{A}]_{\sim_{\mathrm{aff}}}$, and $[\mathcal{A}]_{\sim_{\mathrm{aff}}}$ contains at most four $3$-subsets.

Therefore, sending an affine equivalence class to the union of its $3$-subsets defines a map from $\mathcal{K}'$ to the set of $4$-subsets counted by $N_2$. We show that this map is injective. Let $\mathcal{H}'=\{x,y,z,w\}$ be counted by $N_2$. Then $\tau_{x+y}(z)=x+y+z=w$ by~\eqref{eq004}. Hence, any two $3$-subsets of $\mathcal{H}'$ are affinely equivalent. Therefore, at most one element of $\mathcal{K}'$ is mapped to $\mathcal{H}'$, and we obtain $|\mathcal{K}'|\le N_2$.

Since each element of $\mathcal{K}'$ contains at most four $3$-subsets, we obtain
\begin{equation}\label{eq010}
\binom{n}{3}\le|\mathcal{M}_3(\mathcal{S})|+3|\mathcal{K}'|\le|\mathcal{M}_3(\mathcal{S})|+3N_2 .
\end{equation}

For each $d\in\mathbb{F}_q^*$, now let $\mathcal{V}_d$ be the set of pairs $\{a,a+d\}\subseteq\mathcal{S}$. Then $\sum_{d\ne0}|\mathcal{V}_d|=\binom{n}{2}$.

\begin{claim}\label{claim004}
For every $d\in\mathbb{F}_q^*$, we have $|\mathcal{V}_d|\le2$.
\end{claim}

\textit{Proof of Claim~\ref{claim004}:} Assume, for contradiction, that $\mathcal{V}_d$ contains three distinct pairs $\{a_1,a_1+d\}$, $\{a_2,a_2+d\}$ and $\{a_3,a_3+d\}$. Since $q$ is even, we have $2d=0$. Hence, for every $b\in\mathbb{F}_q$, we have $\{b,b+d\}=\{b+d,(b+d)+d\}$. Therefore, a pair in $\mathcal{V}_d$ is determined by any one of its elements, and two pairs in $\mathcal{V}_d$ with a common element are equal. Consequently, the three pairs are pairwise disjoint.
 
Let $\mathcal{R}':=\{a_1,a_2,a_3\}$. Then $\mathcal{R}'+d=\{a_1+d,a_2+d,a_3+d\}$. Both $\mathcal{R}'$ and $\mathcal{R}'+d$ are $3$-subsets of $\mathcal{S}$, and they are disjoint. In particular, they are distinct. The translation $u\mapsto u+d$ maps $\mathcal{R}'$ onto $\mathcal{R}'+d$. Hence, its restriction to $\mathcal{R}'$ is an affine equivalence from $\mathcal{R}'$ onto $\mathcal{R}'+d$. Since $\mathcal{S}$ is ordering-free, Corollary~\ref{corollary013} and Lemma~\ref{lemma015} show that this affine equivalence exchanges two points of $\mathcal{R}'\cap(\mathcal{R}'+d)$. This is impossible, since $\mathcal{R}'\cap(\mathcal{R}'+d)=\varnothing$. This proves Claim~\ref{claim004}.

Let $\mathcal{H}'=\{x,y,z,w\}$ be counted by $N_2$. Since $x+y+z+w=0$ and $q$ is even, we have $x+y=z+w$. Let $d:=x+y\ne0$. Then $y=x+d$ and $w=z+d$. Hence, $\{x,y\}$ and $\{z,w\}$ are two distinct elements of $\mathcal{V}_d$. The same holds for each of the three partitions of $\mathcal{H}'$ into two pairs. Therefore, each of these partitions, say $\{\{x,y\},\{z,w\}\}$, is a $2$-subset of $\mathcal{V}_d$ for some $d\in\mathbb{F}_q^*$. This $2$-subset determines the partition, and it determines $\mathcal{H}'=\{x,y\}\cup\{z,w\}$. Hence, different choices of $\mathcal{H}'$ and of its partition give distinct $2$-subsets, and we obtain
\[
3N_2\le\sum_{d\ne0}\binom{|\mathcal{V}_d|}{2}.
\]
Since $|\mathcal{V}_d|\le2$ by Claim~\ref{claim004}, we have $\binom{|\mathcal{V}_d|}{2}\le|\mathcal{V}_d|/2$. Since $\sum_{d\ne0}|\mathcal{V}_d|=\binom{n}{2}$, we obtain
\[
3N_2\le\frac12\binom{n}{2}=\frac{n(n-1)}4 .
\]
By~\eqref{eq010}, it follows that $|\mathcal{M}_3(\mathcal{S})|\ge\binom{n}{3}-\frac{n(n-1)}4$.
\end{IEEEproof}

\begin{theorem}\label{theorem021}
Let $\mathcal{S}\subseteq\mathbb{F}_q$ be ordering-free, with $|\mathcal{S}|=n\ge3$.
\begin{enumerate}[(a)]
\item We have $n(n-1)(2n-7)\le2q+16$. In particular, if $n\ge6$, then $n<q^{1/3}+2$.
\item If $q=r^3$ for a prime power $r\ge3$, then $n\le r+1$.
\end{enumerate}
\end{theorem}

\begin{IEEEproof}
(a) By Lemma~\ref{lemma011}(c), the map $[\mathcal{A}]_{\sim_{\mathrm{aff}}}\mapsto\mathcal{O}(\mathcal{A})$ from $\mathcal{M}_3(\mathcal{S})$ to $\mathcal{P}_q$ is injective. By Lemma~\ref{lemma011}(d), we obtain
\begin{equation}\label{eq011}
|\mathcal{M}_3(\mathcal{S})|\le|\mathcal{P}_q|\le\frac{q+8}{6}.
\end{equation}
Combining~\eqref{eq011} with Lemma~\ref{lemma020}, we have
$\frac{n(n-1)(2n-7)}{12}=\binom{n}{3}-\frac{n(n-1)}4\le\frac{q+8}6$, i.e., $n(n-1)(2n-7)\le2q+16$. Now let $n\ge6$, and assume, for contradiction, that $n\ge q^{1/3}+2$. Then $q\le(n-2)^3$, and hence $n(n-1)(2n-7)\le2(n-2)^3+16$, i.e., $3n^2-17n\le0$. This is impossible for $n\ge6$.

(b) Let $\mathcal{S}'\subseteq\mathcal{S}$ with $|\mathcal{S}'|\ge3$. Any ordering of $\mathcal{S}'$ extends to an ordering of $\mathcal{S}$. The increasing triples of the first ordering are increasing triples of the second one. Hence, by Lemma~\ref{lemma007}, the set $\mathcal{S}'$ is also ordering-free.

Assume, for contradiction, that $n\ge r+2$. Then $\mathcal{S}$ has an ordering-free subset $\mathcal{S}'$ with $|\mathcal{S}'|=r+2$. By (a) applied to $\mathcal{S}'$, we have $(r+2)(r+1)(2r-3)\le2r^3+16$, i.e., $3r^2-5r-22\le0$, which is impossible for $r\ge4$.

It remains to consider $r=3$. Then $q=27$ and $|\mathcal{S}'|=5$. By Lemma~\ref{lemma020} and~\eqref{eq011} for $\mathcal{S}'$, we have $5\le|\mathcal{M}_3(\mathcal{S}')|\le\lfloor35/6\rfloor=5$, which yields $|\mathcal{M}_3(\mathcal{S}')|=5$. Then $\binom{5}{3}-|\mathcal{M}_3(\mathcal{S}')|=5$ is odd. This contradicts Lemma~\ref{lemma020}, since $q$ is odd.
\end{IEEEproof}
\begin{remark}\label{remark003}
The assumption $r\ge3$ in Theorem~\ref{theorem021}(b) cannot be removed. For $r=2$, the set in Example~\ref{example019} is an ordering-free subset of $\mathbb{F}_8$ with $r+2=4$ elements.
\end{remark}
Theorem~\ref{theorem021}, Remark~\ref{remark003} and Theorem~\ref{theorem025} answer \textbf{Problem}~\ref{problem002}. For $r\ge3$, an ordering-free evaluation set in $\mathbb{F}_{r^3}$ has at most $r+1$ points. Beyond this length, the ordering must be taken into account, as in Example~\ref{example027}.

\section{Permutation-Robust Evaluation Sets Attaining the Upper Bound}\label{section005}
In this section, we answer \textbf{Problem}~\ref{problem003}. Since ordering-free sets in $\mathbb{F}_{r^3}$ have at most $r+1$ points for $r\ge3$ by Theorem~\ref{theorem021}(b), relaxing permutation robustness to ordering-freeness does not lead to longer evaluation sets. Therefore, we construct permutation-robust sets in $\mathbb{F}_{r^3}$ with $r+1$ points. The construction uses the invariance of cross-ratios under M\"obius transformations.

\subsection{The Projective Line and Cross-Ratios}\label{subsection005001}
For a finite field $\mathbb{F}_q$, let $\mathbb{P}^1(\mathbb{F}_q):=\mathbb{F}_q\cup\{\infty\}$. A nonzero vector $(x,y)\in\mathbb{F}_q^2$ represents the point $x/y$ if $y\ne0$, and the point $\infty$ if $y=0$. For $t\in\mathbb{F}_q$, let $\boldsymbol{\eta}_t:=(t,1)$, and let $\boldsymbol{\eta}_\infty:=(1,0)$. Then $\boldsymbol{\eta}_a$ represents $a$ for every $a\in\mathbb{P}^1(\mathbb{F}_q)$. For $\mathbf{u}=(x,y)$ and $\mathbf{u}'=(x',y')$ in $\mathbb{F}_q^2$, let $[\mathbf{u},\mathbf{u}']:=xy'-x'y$. For $a,b\in\mathbb{P}^1(\mathbb{F}_q)$, let $[a,b]:=[\boldsymbol{\eta}_a,\boldsymbol{\eta}_b]$. The group $\mathrm{PGL}(2,\mathbb{F}_q)$ acts on $\mathbb{P}^1(\mathbb{F}_q)$ by M\"obius transformations~\cite[Equation~(2)]{LY278}. A matrix $\mathbf{M}=\left(\begin{smallmatrix}a_{11}&a_{12}\\a_{21}&a_{22}\end{smallmatrix}\right)\in\mathrm{GL}(2,\mathbb{F}_q)$ sends $(x,y)$ to $(a_{11}x+a_{12}y,\,a_{21}x+a_{22}y)$, and hence gives the M\"obius transformation $t\mapsto(a_{11}t+a_{12})/(a_{21}t+a_{22})$. Here, $t$ is sent to $\infty$ if $a_{21}t+a_{22}=0$, and $\infty$ is sent to $a_{11}/a_{21}$ if $a_{21}\ne0$ and to $\infty$ if $a_{21}=0$. For pairwise distinct points $w,a,b,c\in\mathbb{P}^1(\mathbb{F}_q)$, let
\[
\mathrm{CR}(w,a;b,c):=\frac{[c,a]\,[b,w]}{[b,a]\,[c,w]},
\]
which is the cross-ratio of the four distinct points $w,a,b,c$ in~\cite[Equation~(6)]{LY278}. Since $[s,t]=s-t$ and $[t,\infty]=-1$ for $s,t\in\mathbb{F}_q$, for pairwise distinct $a,b,c\in\mathbb{F}_q$ we have
\begin{equation}\label{eq012}
\rho(a,b,c)=\mathrm{CR}(\infty,a;b,c).
\end{equation}
The next lemma collects three known facts on $\mathrm{PGL}(2,\mathbb{F}_q)$ that are used in the proof of Lemma~\ref{lemma024}.
\begin{lemma}\label{lemma022}
Let $q$ be a prime power.
\begin{enumerate}[(a)]
\item The group $\mathrm{PGL}(2,\mathbb{F}_q)$ acts sharply $3$-transitively on $\mathbb{P}^1(\mathbb{F}_q)$~\cite[Theorem~3.1(i)]{LY278}.
\item Let $w,a,b,c$ be four distinct points of $\mathbb{P}^1(\mathbb{F}_q)$. Then $\mathrm{CR}(\pi(w),\pi(a);\pi(b),\pi(c))=\mathrm{CR}(w,a;b,c)$ for every $\pi\in\mathrm{PGL}(2,\mathbb{F}_q)$~\cite[Theorem~3.1(ii)]{LY278}.
\item The group $\mathrm{PGL}(2,\mathbb{F}_q)$ is a subgroup of $\mathrm{PGL}(2,\mathbb{F}_{q^3})$, and it acts regularly on $\mathbb{P}^1(\mathbb{F}_{q^3})\setminus\mathbb{P}^1(\mathbb{F}_q)=\mathbb{F}_{q^3}\setminus\mathbb{F}_q$, see the remark following~\cite[Theorem~6]{LY315}. In particular, for every $\theta\in\mathbb{F}_{q^3}\setminus\mathbb{F}_q$, the map $\pi\mapsto\pi(\theta)$ is a bijection from $\mathrm{PGL}(2,\mathbb{F}_q)$ onto $\mathbb{F}_{q^3}\setminus\mathbb{F}_q$~\cite[Chapter~I, \S5, no.~6]{LYBOOK030}.
\end{enumerate}
\end{lemma}
\subsection{Construction and Permutation-Robustness}
Throughout this subsection, let $r$ be a prime power.
\begin{definition}\label{definition023}
Let $\theta\in \mathbb{F}_{r^3}\setminus \mathbb{F}_r$. Let $\psi(t):=(t-\theta)^{-1}$ for $t\in \mathbb{F}_r$ and $\psi(\infty):=0$, and let
\[
\mathcal{X}_\theta:=\psi(\mathbb{P}^1(\mathbb{F}_r))=\{(t-\theta)^{-1}:t\in \mathbb{F}_r\}\cup\{0\}.
\]
\end{definition}
We extend $\psi$ to $\mathbb{P}^1(\mathbb{F}_{r^3})$ so that it sends $\theta$ to $\infty$. Then, by Lemma~\ref{lemma022}(b) and~\eqref{eq012}, the ratios of the points of $\mathcal{X}_\theta$ are cross-ratios of points of $\mathbb{P}^1(\mathbb{F}_r)$ with $\theta$. The next lemma makes this precise.
\begin{lemma}\label{lemma024}
Let $\theta\in \mathbb{F}_{r^3}\setminus \mathbb{F}_r$, and extend $\psi$ to $\mathbb{P}^1(\mathbb{F}_{r^3})$ by $\psi(s)=(s-\theta)^{-1}$ for $s\in \mathbb{F}_{r^3}\setminus\{\theta\}$, $\psi(\theta)=\infty$ and $\psi(\infty)=0$.
\begin{enumerate}[(a)]
\item The map $\psi$ lies in $\mathrm{PGL}(2,\mathbb{F}_{r^3})$. In particular, $\psi$ is a bijection of $\mathbb{P}^1(\mathbb{F}_{r^3})$.
\item For pairwise distinct $a,b,c\in\mathbb{P}^1(\mathbb{F}_{r^3})\setminus\{\theta\}$, the points $\psi(a),\psi(b),\psi(c)$ are pairwise distinct elements of $\mathbb{F}_{r^3}$, and $\rho(\psi(a),\psi(b),\psi(c))=\mathrm{CR}(\theta,a;b,c)$.
\item The map $(a,b,c)\mapsto\rho(\psi(a),\psi(b),\psi(c))$ is a bijection from $\mathcal{T}(\mathbb{P}^1(\mathbb{F}_r))$ onto $\mathbb{F}_{r^3}\setminus \mathbb{F}_r$.
\end{enumerate}
\end{lemma}
\begin{IEEEproof}
(a) Let $\mathbf{M}_\psi:=\left(\begin{smallmatrix}0&1\\1&-\theta\end{smallmatrix}\right)$. Since $\det\mathbf{M}_\psi=-1$, we have $\mathbf{M}_\psi\in\mathrm{GL}(2,\mathbb{F}_{r^3})$. The M\"obius transformation given by $\mathbf{M}_\psi$ is $s\mapsto(0\cdot s+1)/(s-\theta)=(s-\theta)^{-1}$ for $s\in\mathbb{F}_{r^3}\setminus\{\theta\}$, and by the conventions above it sends $\theta$ to $\infty$ and $\infty$ to $0$. Hence, it coincides with $\psi$. In particular, $\psi\in\mathrm{PGL}(2,\mathbb{F}_{r^3})$ is a bijection of $\mathbb{P}^1(\mathbb{F}_{r^3})$.

(b) Let $a,b,c\in\mathbb{P}^1(\mathbb{F}_{r^3})\setminus\{\theta\}$ be pairwise distinct. Since $\psi$ is injective by (a) and $\psi(\theta)=\infty$, the points $\psi(a),\psi(b),\psi(c)$ are pairwise distinct elements of $\mathbb{F}_{r^3}$. By~\eqref{eq012} and Lemma~\ref{lemma022}(b) with $q=r^3$ and $\pi=\psi$, we obtain
\[
\rho(\psi(a),\psi(b),\psi(c))=\mathrm{CR}\bigl(\infty,\psi(a);\psi(b),\psi(c)\bigr)=\mathrm{CR}\bigl(\psi(\theta),\psi(a);\psi(b),\psi(c)\bigr)=\mathrm{CR}(\theta,a;b,c).
\]

(c) Let $(a,b,c)\in\mathcal{T}(\mathbb{P}^1(\mathbb{F}_r))$. On the one hand, by Lemma~\ref{lemma022}(a), there is a unique $\pi_{abc}\in\mathrm{PGL}(2,\mathbb{F}_r)$ with $\pi_{abc}(a)=1$, $\pi_{abc}(b)=0$ and $\pi_{abc}(c)=\infty$, and the map $(a,b,c)\mapsto\pi_{abc}$ is a bijection from $\mathcal{T}(\mathbb{P}^1(\mathbb{F}_r))$ onto $\mathrm{PGL}(2,\mathbb{F}_r)$.

On the other hand, by Lemma~\ref{lemma022}(c) with $q=r$, the map $\pi\mapsto\pi(\theta)$ is a bijection from $\mathrm{PGL}(2,\mathbb{F}_r)$ onto $\mathbb{F}_{r^3}\setminus\mathbb{F}_r$.

Since $\pi_{abc}\in\mathrm{PGL}(2,\mathbb{F}_{r^3})$ by Lemma~\ref{lemma022}(c), we obtain from (b) and Lemma~\ref{lemma022}(b) with $q=r^3$ that
\begin{equation}\label{eq013}
\begin{aligned}
\rho(\psi(a),\psi(b),\psi(c))&=\mathrm{CR}(\theta,a;b,c)=\mathrm{CR}\bigl(\pi_{abc}(\theta),1;0,\infty\bigr)\\
&=\frac{[\infty,1]\,[0,\pi_{abc}(\theta)]}{[0,1]\,[\infty,\pi_{abc}(\theta)]}=\frac{1\cdot(-\pi_{abc}(\theta))}{(-1)\cdot1}=\pi_{abc}(\theta).
\end{aligned}
\end{equation}
Composing the bijections $(a,b,c)\mapsto\pi_{abc}$ and $\pi\mapsto\pi(\theta)$, we obtain that the map $(a,b,c)\mapsto\pi_{abc}(\theta)$ is a bijection from $\mathcal{T}(\mathbb{P}^1(\mathbb{F}_r))$ onto $\mathbb{F}_{r^3}\setminus\mathbb{F}_r$. By~\eqref{eq013}, this map is $(a,b,c)\mapsto\rho(\psi(a),\psi(b),\psi(c))$, which completes the proof.
\end{IEEEproof}

\begin{theorem}\label{theorem025}
Let $\theta\in \mathbb{F}_{r^3}\setminus \mathbb{F}_r$. Then $\mathcal{X}_\theta$ is permutation-robust with $r+1$ points. Consequently, every ordering of $\mathcal{X}_\theta$ gives an ${[r+1,2]}_{r^3}$ RS code attaining the half-Singleton bound.
\end{theorem}
\begin{IEEEproof}
By Lemma~\ref{lemma024}(a), the map $\psi$ is injective. Hence, $\psi$ maps $\mathbb{P}^1(\mathbb{F}_r)$ bijectively onto $\mathcal{X}_\theta=\psi(\mathbb{P}^1(\mathbb{F}_r))$, and $\mathcal{X}_\theta\subseteq\mathbb{F}_{r^3}$ by Definition~\ref{definition023}. Since $|\mathbb{P}^1(\mathbb{F}_r)|=r+1$, we obtain $|\mathcal{X}_\theta|=r+1$. Since $\psi$ maps $\mathbb{P}^1(\mathbb{F}_r)$ bijectively onto $\mathcal{X}_\theta$, the map $(a,b,c)\mapsto(\psi(a),\psi(b),\psi(c))$ is a bijection from $\mathcal{T}(\mathbb{P}^1(\mathbb{F}_r))$ onto $\mathcal{T}(\mathcal{X}_\theta)$. By Lemma~\ref{lemma024}(c), the map $(a,b,c)\mapsto\rho(\psi(a),\psi(b),\psi(c))$ is injective on $\mathcal{T}(\mathbb{P}^1(\mathbb{F}_r))$. Therefore, the ratios of the elements of $\mathcal{T}(\mathcal{X}_\theta)$ are pairwise distinct, i.e., $\mathcal{X}_\theta$ is permutation-robust. Since $r$ is a prime power, we have $r\ge2$, and hence $|\mathcal{X}_\theta|=r+1\ge3$. By Lemma~\ref{lemma017}(b), $\mathcal{X}_\theta$ is ordering-free, i.e., every ordering of $\mathcal{X}_\theta$ gives an ${[r+1,2]}_{r^3}$ RS code attaining the half-Singleton bound.
\end{IEEEproof}
By Proposition~\ref{proposition018}, the size $r+1$ of $\mathcal{X}_\theta$ is the largest possible for a permutation-robust subset of $\mathbb{F}_{r^3}$. For $r\ge3$, it is also the largest possible for an ordering-free subset by Theorem~\ref{theorem021}(b). Hence, Theorem~\ref{theorem025} answers \textbf{Problem}~\ref{problem003}. 
\begin{remark}\label{remark005}
Let $\mathcal{S}\subseteq\mathcal{X}_\theta$ with $|\mathcal{S}|\ge3$. Since $\mathcal{T}(\mathcal{S})\subseteq\mathcal{T}(\mathcal{X}_\theta)$, the set $\mathcal{S}$ is also permutation-robust, and hence ordering-free by Lemma~\ref{lemma017}(b). Therefore, for every $3\le n\le r+1$, every ordering of an $n$-subset of $\mathcal{X}_\theta$ gives an optimal ${[n,2]}_{r^3}$ RS code.
\end{remark}
\begin{remark}\label{remark006}
By Theorem~\ref{theorem021}(a), an ordering-free set in $\mathbb{F}_q$ with $n$ points requires $q\ge n(n-1)(2n-7)/2-8$. If $n-1$ is a prime power, then $\mathcal{X}_\theta$ with $r=n-1$ is an ordering-free set with $n$ points in $\mathbb{F}_q$ for $q=(n-1)^3$. Hence, if $n-1$ is a prime power, then the smallest field size for ordering-free sets with $n$ points is $n^3(1+o(1))$. In contrast, for a single ordering, Corollary~\ref{corollary008} only gives $q\ge\binom{n}{3}+2$, which is about $n^3/6$.
\end{remark}    
\begin{corollary}\label{corollary026}
Let $\theta\in\mathbb{F}_{r^3}\setminus\mathbb{F}_r$, and let $c\in\mathbb{F}_{r^3}$ with $0\notin\mathcal{X}_\theta+c$. Then $\mathcal{X}_\theta+c$ gives a fully anonymous $(1,3,r+1)$ ramp secret sharing scheme over $\mathbb{F}_{r^3}$ in the sense of~\cite{LY46}.
\end{corollary}
\begin{IEEEproof}
Since translations preserve ratios by~\eqref{eq001}, $\mathcal{X}_\theta+c$ is permutation-robust by Theorem~\ref{theorem025}. By~\cite[Claim~1]{LY46} and Lemma~\ref{lemma011}(b), a two-dimensional RS code of length $n$ is robust against the $(n-3)$-permutation-insdel adversary of~\cite{LY46} if and only if its evaluation set is permutation-robust. Hence, the construction in the proof of~\cite[Corollary~1]{LY46} with $k=2$, applied to $\mathcal{X}_\theta+c$, gives the required scheme. 
\end{IEEEproof}
Over the same field $\mathbb{F}_{r^3}$, the scheme in Corollary~\ref{corollary026} has $r+1$ participants, while the scheme in~\cite[Corollary~2]{LY46} has at most $r-1$ participants.
\begin{example}\label{example027}
Let $\mathbb{F}_{27}=\mathbb{F}_3[\theta]/(\theta^3+2\theta^2+1)$. By Definition~\ref{definition023}, we have $\mathcal{X}_\theta=\{0,\theta^2,2\theta+\theta^2,1+2\theta+2\theta^2\}$, and its $24$ ratios on ordered triples are exactly the elements of $\mathbb{F}_{27}\setminus\mathbb{F}_3$, as stated in Lemma~\ref{lemma024}(c). Hence, every ordering of $\mathcal{X}_\theta$ gives an optimal ${[4,2]}_{27}$ RS code. On the other hand, let $\boldsymbol{\alpha}=(0,\,1,\,2\theta^2,\,1+\theta,\,\theta,\,2\theta)$. Its $20$ ratios $\rho_{ijk}$, listed in Table~\ref{table002}, are pairwise distinct. By Lemma~\ref{lemma007}, $\boldsymbol{\alpha}$ gives an optimal ${[6,2]}_{27}$ RS code. By Theorem~\ref{theorem021}(b), the evaluation set of $\boldsymbol{\alpha}$ is not ordering-free. For example, the ordering $(0,1,2\theta^2,1+\theta,2\theta,\theta)$ of this set is not optimal, since $\rho_{234}=\rho_{356}$.
\end{example}
\begin{table}[ht]\centering\small
\caption{The Ratios $\rho_{ijk}$ of the ${[6,2]}_{27}$ Code of Example~\ref{example027}, in the Basis $1,\theta,\theta^2$}\label{table002}
\begin{tabular}{lllll}\toprule
$\rho_{123}=2\theta^2$ & $\rho_{124}=1+\theta$ & $\rho_{125}=\theta$ & $\rho_{126}=2\theta$ & $\rho_{134}=2+\theta^2$\\
$\rho_{135}=2\theta+\theta^2$ & $\rho_{136}=\theta+2\theta^2$ & $\rho_{145}=2+2\theta+2\theta^2$ & $\rho_{146}=1+\theta+\theta^2$ & $\rho_{156}=2$\\
$\rho_{234}=1+\theta^2$ & $\rho_{235}=1+\theta+2\theta^2$ & $\rho_{236}=2+\theta$ & $\rho_{245}=1+2\theta+\theta^2$ & $\rho_{246}=2+2\theta+\theta^2$\\
$\rho_{256}=2+2\theta^2$ & $\rho_{345}=2+\theta+2\theta^2$ & $\rho_{346}=1+2\theta^2$ & $\rho_{356}=\theta+\theta^2$ & $\rho_{456}=1+2\theta$\\\bottomrule
\end{tabular}
\end{table}

\section{Concluding Remarks and Further Research}\label{section006}
In this paper, we studied two-dimensional RS codes that attain the half-Singleton bound, with emphasis on how their minimum insdel distances depend on the ratios and on the ordering of their evaluation points.

We first showed that for $n\ge6$ every two-dimensional linear insdel code attaining the half-Singleton bound is an RS code (Theorem~\ref{theorem009}), which answers \textbf{Problem}~\ref{problem001}. As a consequence, every such code needs $q\ge\binom{n}{3}+2$, and this answers the case $k=2$ and $n\ge6$ of the question in~\cite[Section~1.4]{LY104}. \textbf{Problem}~\ref{problem002} concerns the largest size of an ordering-free set in $\mathbb{F}_q$. An ordering-free set in $\mathbb{F}_q$ with $n\ge6$ points satisfies $n<q^{1/3}+2$ (Theorem~\ref{theorem021}(a)). Over $\mathbb{F}_{r^3}$, permutation-robust sets have at most $r+1$ points (Proposition~\ref{proposition018}), and for $r\ge3$ the same bound holds for ordering-free sets (Theorem~\ref{theorem021}(b)). The condition $r\ge3$ cannot be removed (Remark~\ref{remark003}). Finally, for every prime power $r$, the sets $\mathcal{X}_\theta$ are permutation-robust with $r+1$ points (Theorem~\ref{theorem025}). They attain the bound of Proposition~\ref{proposition018}, and also that of Theorem~\ref{theorem021}(b) for $r\ge3$. This answers \textbf{Problem}~\ref{problem003}. As an application, translates of the sets $\mathcal{X}_\theta$ give fully anonymous $(1,3,r+1)$ ramp secret sharing schemes over $\mathbb{F}_{r^3}$ (Corollary~\ref{corollary026}).

A natural direction for future research is to give systematic constructions of ${[n,2]}_{r^3}$ RS codes attaining the half-Singleton bound with $n\ge r+2$. By Theorem~\ref{theorem021}(b), for $r\ge3$ such a code attains this bound only for suitable orderings of its evaluation set, as in Example~\ref{example027}. Hence, these constructions must take the coordinate ordering into account.

\section*{Acknowledgment}
During the preparation of this manuscript, the authors used Claude Opus 5.5 for English-language editing, numerical verification of the examples, and preliminary checks of proofs. The authors take full responsibility for the mathematical statements, proofs, originality claims, and references in the manuscript.

\end{document}